\documentclass[twocolumn,english,pra]{revtex4}
\usepackage[T1]{fontenc}
\usepackage[latin9]{inputenc}
\usepackage{color}
\usepackage{amsmath}
\usepackage{amssymb}
\usepackage{graphicx}
\usepackage{esint}

\makeatletter
\@ifundefined{textcolor}{}
{%
 \definecolor{BLACK}{gray}{0}
 \definecolor{WHITE}{gray}{1}
 \definecolor{RED}{rgb}{1,0,0}
 \definecolor{GREEN}{rgb}{0,1,0}
 \definecolor{BLUE}{rgb}{0,0,1}
 \definecolor{CYAN}{cmyk}{1,0,0,0}
 \definecolor{MAGENTA}{cmyk}{0,1,0,0}
 \definecolor{YELLOW}{cmyk}{0,0,1,0}
}

\usepackage{babel}

\makeatother

\usepackage{babel}
\begin{document}

\title{EPR paradox and quantum steering in a three-mode optomechanical system}

\author{Qiongyi He$^{1,2}$}

\email{qiongyihe@pku.edu.cn}

\author{Zbigniew Ficek$^{3}$}

\affiliation{$^{1}$State Key Laboratory of Mesoscopic Physics, Department of
Physics, Peking University, Beijing 100871, China}

\affiliation{$^{2}$Collaborative Innovation Center of Quantum Matter, Beijing
100871, China}

\affiliation{$^{3}$The National Centre for Mathematics and Physics, KACST, P.O.
Box 6086, Riyadh 11442, Saudi Arabia}
\begin{abstract}
We study multi-partite entanglement, the generation of EPR states
and quantum steering in a three-mode optomechanical system composed
of an atomic ensemble located inside a single-mode cavity with a movable
mirror. The cavity mode is driven by a short laser pulse, has a nonlinear
parametric-type interaction with the mirror and a linear beamsplitter-type
interaction with the atomic ensemble. There is no direct interaction
of the mirror with the atomic ensemble. A threshold effect for the
dynamics of the system is found, above which the system works as an
amplifier and below which as an attenuator of the output fields. The
threshold is determined by the ratio of the coupling strengths of
the cavity mode to the mirror and to the atomic ensemble. It is shown
that above the threshold the system effectively behaves as a two-mode
system in which a perfect bipartite EPR state can be generated, while
it is impossible below the threshold. Furthermore, a fully inseparable
tripartite entanglement and even further a genuine tripartite entanglement
can be produced above and below the threshold. In addition, we consider
quantum steering and examine the monogamy relations that quantify
the amount of bipartite steering that can be shared between different
modes. It is found that the mirror is more capable for steering of
entanglement than the cavity mode. The two way steering is found between
the mirror and the atomic ensemble despite the fact that they are
not directly coupled to each other, while it is impossible between
the output of cavity mode and the ensemble which are directly coupled
to each other. 
\end{abstract}


\maketitle

\section{Introduction}

The possibility of entangling macroscopic objects has been of interest
for many years. Of particular interest is the possibility to entangle
macroscopic objects using an optomechanical cavity containing a macroscopic
mechanical oscillator such as a movable mirror or vibrating membrane
\cite{gm09,agh10,td09,gh09}. In an optomechanical system the motion
of the mechanical oscillator can be affected by the radiation pressure
of the cavity field which may result in a parametric coupling between
the cavity mode and the oscillator. The parametric couplings have
long been known to produce nonclassical effects such as multimode
squeezing and entanglement. Consequently, a number of papers have
been devoted to the study of entanglement in optomechanical systems.
Most of these studies considers two-mode optomechanical systems and
explores entanglement in the steady-state \cite{fg06,vg07,pv07,vt07,bg08,gm08,ig08,pc10,cp11,mg02,hp08,hw09,wh10,ck11,jh09,bv11,zh11,tb13}.
Further studies have considered the generation of steady-state bipartite
entanglement in three-mode systems \cite{gv08,sl12,wc13}.

In contrast to the numerous publications concerning the steady-state
entanglement there are only a limited number of studies in transient
(pulsed) regime. In terms of the difficulties in the creation of stationary
entanglement, the pulsed regime could be free from the decoherence
and dissipation effects such as the damping of the oscillating mirror.
In addition, it could not be limited by the stability conditions imposed
on the steady-state solutions. Pulsed excitation of a two-mode optomechanical
system has been treated by several authors in the context of the generation
of EPR-type correlations between the pulse and a mechanical oscillator
\cite{hw11} and quantum steering \cite{hr13}. When an optomechanical
system is composed of more than two modes, more complex correlations
can be created. These correlations can significantly affect the two-mode
entanglement and result in multimode entanglement.

It is the purpose of this paper to study bipartite and tripartite
entanglement in a three-mode system realized with an atomic ensemble
located inside a single-mode cavity with a movable mirror. We work
in the pulsed regime and concentrate on the ability of the system
to generate perfect bipartite EPR states, fully inseparable and genuine
tripartite entanglement and quantum steering. We assume that in the
pulsed regime the relaxation effects of the mirror and the atoms can
be neglected. In fact, it is not an overly restrictive limitation
regarding a slow damping rate of the mirror and the existence of dipole
transitions with relatively long relaxation times in some atoms~\cite{by85}.
We find a threshold effect for the dynamics of the system imposed
by the ratio $G/G_{a}$ of the coupling strengths of the oscillating
mirror and the atoms to the cavity mode, respectively. Above the threshold
$(G>G_{a})$, the system behaves as an amplifier, whereas below the
threshold $(G<G_{a})$, the system behaves as an attenuator of the
input laser pulses. The threshold behave leads to substantially different
results for the bipartite and tripartite entanglement above and below
the threshold.

The paper is organized as follows. In Sec. \ref{sec:2} we develop
a general formalism for the pulsed three-mode optomechanics. The formalism
is based on solution the Heisenberg equations of motion, which is
an extension of the work of Hofer \textit{et al.} \cite{hw11} to
the case of three coupled modes. General expressions for the input-output
relations between the quadrature components of the modes are derived
and some of their properties are discussed along with a brief discussion
of the variances of the quadrature components in the limit of large
squeezing. Section \ref{sec:3} is concerned with bipartite entanglement
between the modes. We adopt both the symmetric criterion of Duan,
Giedke, Cirac, and Zoller (DGCZ) \cite{DGCZ} and a less restrictive
condition, based on asymmetric weightings of the quadratures \cite{hr13,Simon,Tombesi}
to examine the bipartite entanglement and the inseparability conditions
as a function of the duration of the input laser pulse. Explicit analytic
solutions are given for the parameters and the conditions for entanglement
and the possibility to generate perfect EPR entangled states are determined.
Section \ref{sec:4} is devoted to tripartite entanglement. In searching
for fully inseparable tripartite entanglement, we use the van Loock-Furusawa
inequalities \cite{van03}, and for the detection of genuine tripartite
entanglement we use the criterion given by Shalm et al \cite{genuineNP2012}
and Reid \cite{genuineentM}. In Sec. \ref{sec:5} we concentrate
on quantum steering and examine the monogamy relations that quantify
the amount of bipartite steering that can be shared between different
modes. We consider the monogamy relations and inequalities for quantum
steering, recently proposed by Reid \cite{steerM}. We summarize our
results in Sec. \ref{sec:6}. Finally, in the Appendix, we give the
explicit expressions for the input-output relations between the quadrature
components of the three modes of the system.

\section{Pulsed three-mode optomechanics\label{sec:2}}

\begin{figure}[t]
\centering{}\includegraphics[width=0.85\columnwidth]{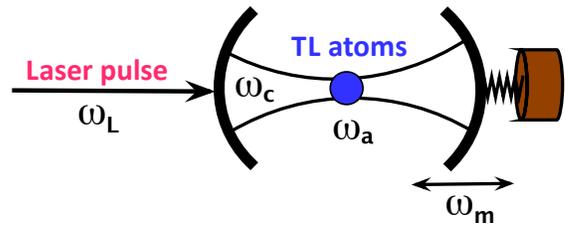} \caption{(Color online) Schematic diagram of the system. An ensemble of identical
two-level (TL) atoms, each of transition frequency $\omega_{a}$,
is located inside a single-mode cavity of frequency $\omega_{c}$.
The cavity mode is driven by a laser pulse of frequency$\omega_{L}$
and the movable mirror oscillates with frequency $\omega_{m}$.\label{fig:1} }
\end{figure}

We consider an optomechanical system composed of an ensemble of $N$
identical two-level atoms located inside a single-mode cavity formed
by two mirrors, a fixed semitransparent mirror and a movable fully
reflective mirror, as shown in Fig. \ref{fig:1}. The cavity mode
is driven through the semitransparent mirror by a pulsed laser field,
which is treated classically in our calculations and is characterized
by its frequency $\omega_{L}$ and a time dependent amplitude $E(t)$.
We assume that the amplitude $E(t)$ is constant over a short time
interval $0\leq t\leq\tau$ and zero outside this interval. The cavity
field is treated as quantized and is characterized by its frequency
$\omega_{c}$ and the annihilation and creation operators $a_{c}$
and $a_{c}^{\dagger}$. The movable mirror is modeled as a quantized
single-mode harmonic oscillator of frequency $\omega_{m}$ and an
amplitude determined by operators $a_{m}$ and $a_{m}^{\dagger}$.
The oscillations of the movable mirror result from the radiation pressure
of the cavity field on the mirror. The atomic ensemble of identical
two-level atoms, each composed of a ground state $|g_{j}\rangle$
and an excited state $|e_{j}\rangle$, separated by the transition
frequency $\omega_{a}$, interacts with the cavity mode and is represented
by the collective dipole lowering, raising, and population difference
operators, ${\color{red}S^{-}=\sum_{j}|g_{j}\rangle\langle e_{j}|}$,
$S^{+}=\sum_{j}|e_{j}\rangle\langle g_{j}|$, and $S_{z}=\sum_{j}(|e_{j}\rangle|e_{j}\rangle-|g_{j}\rangle|g_{j}\rangle)$,
respectively. We assume that the coupling of the atoms to the cavity
mode is weak so that the excitation probability of the atomic ensemble
is very small, $S_{z}\approx\langle S_{z}\rangle\approx-N$. In this
case, we may represent the collective dipole lowering and raising
operators in terms of bosonic annihilation and creation operators,
$c_{a}=S^{-}/\sqrt{|\langle S_{z}\rangle|}$ and $c_{a}^{\dagger}=S^{+}/\sqrt{|\langle S_{z}\rangle|}$,
respectively. It is easily verified that the operators $c_{a}$ and
$c_{a}^{\dagger}$ satisfy the fundamental commutation relation for
boson operators, $[c_{a},c_{a}^{\dagger}]=1$.

The total Hamiltonian $H$ of the system, including the laser driving
field and the nonlinear radiation pressure interaction between the
cavity mode and the mirror, can be written as 
\begin{equation}
H=H_{c}+H_{a}+H_{m}+H_{I},\label{q1}
\end{equation}
where 
\begin{equation}
H_{c}=\hbar\omega_{c}a_{c}^{\dagger}a_{c}\label{q2}
\end{equation}
is the free Hamiltonian of the cavity mode of frequency $\omega_{c}$,
\begin{equation}
H_{a}=\hbar\omega_{a}c_{a}^{\dagger}c_{a}\label{q3}
\end{equation}
is the free Hamiltonian of the atomic excitation mode of frequency$\omega_{a}$,
\begin{equation}
H_{m}=\hbar\omega_{m}a_{m}^{\dagger}a_{m}\label{q4}
\end{equation}
is the free Hamiltonian of the movable mirror oscillating with frequency
$\omega_{m}$, and 
\begin{align}
H_{I} & =\hbar g_{a}\left(c_{a}^{\dagger}a_{c}+a_{c}^{\dagger}c_{a}\right)+\hbar g_{0}a_{c}^{\dagger}a_{c}\left(a_{m}^{\dagger}+a_{m}\right)\nonumber \\
 & \ \ \ \ +i\hbar\left[E(t)a_{c}^{\dagger}{\rm e}^{-i\omega_{L}t}-E^{\ast}(t)a_{c}{\rm e}^{i\omega_{L}t}\right]\label{q5}
\end{align}
is the interaction Hamiltonian of the cavity mode with the atomic
mode, the external driving field, and with the movable mirror. Here,
$g_{a}$ is the coupling constant between the atoms and the cavity
mode, $g_{0}$ is the coupling constant between the cavity mode and
the movable mirror, and $E$ is the amplitude of the laser field.
Note that the atomic mode is not directly coupled to the mechanical
mode and the interaction between the cavity and mechanical modes involves
the nonlinear optomechanical coupling \cite{gv08,hw11}.

To remove the terms oscillating with the laser frequency $\omega_{L}$
in Eq. (\ref{q5}), we introduce the evolution operator 
\begin{equation}
U(t)={\rm e}^{i\omega_{L}(a_{c}^{\dagger}a_{c}+c_{a}^{\dagger}c_{a})t},\label{q6}
\end{equation}
which transforms the Hamiltonian (\ref{q1}) into 
\begin{align}
\tilde{H} & \equiv U(t)\left(H-i\hbar\frac{d}{dt}\right)U^{\dagger}(t)\nonumber \\
 & =\hbar\Delta_{c}a_{c}^{\dagger}a_{c}+\hbar\Delta_{a}c_{a}^{\dagger}c_{a}+\hbar\omega_{m}a_{m}^{\dagger}a_{m}\nonumber \\
 & \ \ \ +\hbar g_{a}\left(c_{a}^{\dagger}a_{c}+a_{c}^{\dagger}c_{a}\right)+\hbar g_{0}a_{c}^{\dagger}a_{c}\left(a_{m}^{\dagger}+a_{m}\right)\nonumber \\
 & \ \ \ +i\hbar\left[E(t)a_{c}^{\dagger}-E^{\ast}(t)a_{c}\right],\label{q7}
\end{align}
where $\Delta_{c}=\omega_{c}-\omega_{L}$ and $\Delta_{a}=\omega_{a}-\omega_{L}$
are the detunings of the atomic frequency $\omega_{c}$ and the cavity
mode frequency $\omega_{a}$, respectively, from the the driving field
frequency $\omega_{L}$. Note that in practice, the frequency $\omega_{m}$
is much smaller than the laser, atomic and cavity frequencies and,
therefore, can be comparable to the detunings $\Delta_{c}$ and $\Delta_{a}$.

The Hamiltonian (\ref{q7}) involves the nonlinear optomechanical
coupling between the cavity mode and the oscillating mirror, which
results from the radiation pressure of the cavity field on the mirror.
In the limit of a strong driving, $|E_{L}|\gg g_{0},g_{a}$, the Hamiltonian
(\ref{q7}) can be significantly simplified and the driving term eliminated
by going into a displaced picture, in which each operator is written
as the sum of its steady-state value and a small linear displacement
\begin{equation}
c_{a}\rightarrow\alpha_{a}+\delta c_{a},\ a_{m}\rightarrow\alpha_{m}+\delta a_{m},\ a_{c}\rightarrow\alpha_{c}+\delta a_{c}.\label{q8}
\end{equation}
The displacement operators satisfy the Gaussian statistics and are
delta correlated in time such that all the first moments vanish, 
\begin{equation}
\langle\delta c_{a}\rangle=\langle\delta a_{m}\rangle=\langle\delta a_{c}\rangle=0,\label{q9}
\end{equation}
and for the second moments we suppose that the only nonzero are 
\begin{align}
\langle\delta c_{a}(t)\delta c_{a}^{\dagger}(t^{\prime})\rangle & =\langle\delta a_{c}(t)\delta a_{c}^{\dagger}(t^{\prime})\rangle=\delta(t-t^{\prime}),\nonumber \\
\langle\delta a_{m}(t)\delta a_{m}^{\dagger}(t^{\prime})\rangle & =(n_{0}+1)\delta(t-t^{\prime}),\nonumber \\
\langle\delta a_{m}^{\dagger}(t)\delta a_{m}(t^{\prime})\rangle & =n_{0}\delta(t-t^{\prime}),\label{q10}
\end{align}
where $n_{0}=[\exp(\hbar\omega_{m}/k_{B}T)-1]^{-1}$ is the mean number
of the thermal photons at the frequency of the mechanical mode, $k_{B}$
is the Boltzmann constant and $T$ is the temperature of the environment
surrounding the mirror. In other words, we assume that the cavity
mode and the atom are in the ordinary zero temperature environment,
whereas the mirror is in a thermal field of a nonzero temperature.

Taking only the quadratic terms of the displacement operators, we
arrive to an effective Hamiltonian 
\begin{align}
H_{eff} & =\hbar\Delta_{a}\delta c_{a}^{\dagger}\delta c_{a}+\hbar\omega_{m}\delta a_{m}^{\dagger}\delta a_{m}+\hbar\Delta_{c}^{\prime}\delta a_{c}^{\dagger}\delta a_{c}\nonumber \\
 & \ \ +\hbar g_{a}\left(\delta c_{a}^{\dagger}\delta a_{c}+\delta a_{c}^{\dagger}\delta c_{a}\right)\nonumber \\
 & \ \ +\hbar g\left(\delta a_{c}+\delta a_{c}^{\dagger}\right)(\delta a_{m}+\delta a_{m}^{\dagger}),\label{q11}
\end{align}
where $\Delta_{c}^{\prime}=\Delta_{c}+g_{0}(\alpha_{m}+\alpha_{m}^{\ast})$
and $g=g_{0}|\alpha_{c}|$.

The Hamiltonians (\ref{q11}) leads to the following Heisenberg equations
of motion for the displacement operators 
\begin{align}
\delta\dot{a}_{m}= & -(\gamma_{m}+i\omega_{m})\delta a_{m}-ig(\delta a_{c}+\delta a_{c}^{\dagger})-\sqrt{2\gamma_{m}}\xi_{{\rm in}},\nonumber \\
\delta\dot{a}_{c}= & -(\kappa+i\Delta_{c}^{\prime})\delta a_{c}-ig_{a}\delta c_{a}\nonumber \\
 & \ \ \ \ \ \ \ \ -ig(\delta a_{m}^{\dagger}+\delta a_{m})-\sqrt{2\kappa}a_{{\rm in}},\nonumber \\
\delta\dot{c}_{a}= & -(\gamma_{a}+i\Delta_{a})\delta c_{a}-ig_{a}\delta a_{c}-\sqrt{2\gamma_{a}}c_{{\rm in}},\label{q12}
\end{align}
where we have included damping rates of the modes and the corresponding
noise operators.

Our purpose of this paper is to examine entangled properties of the
three modes and quantum steering. It is well known that these properties
are strongly sensitive to losses. For this reason we shall work in
the bad cavity limit of $\kappa\gg g_{a},g$ and consider a short
evolution time of the system determined by the damping rate of the
cavity mode, $t\sim1/\kappa$. It has an advantage that over the short
evolution time $t\sim1/\kappa$, the relaxations of the atom and the
mechanical mirror can be neglected $(\gamma_{a}=\gamma_{m}=0)$ together
with the corresponding noise terms, $\xi_{{\rm in}}=c_{{\rm in}}=0$.

In order to solve Eq. (\ref{q12}), it is convenient to introduce
slowly varying variables which are free from the oscillations at the
frequency $\omega_{m}$ and are related to to the annihilation operators
by 
\begin{align}
\delta a_{m}^{r} & =\delta a_{m}{\rm e}^{i\omega_{m}t},\quad\delta a_{c}^{r}=\delta a_{c}{\rm e}^{-i\omega_{m}t},\nonumber \\
\delta c_{a}^{r} & =\delta c_{a}{\rm e}^{-i\omega_{m}t},\quad a_{{\rm in}}^{r}=a_{{\rm in}}{\rm e}^{-i\omega_{m}t}.\label{q13}
\end{align}
In terms of these new variables and after dropping the damping and
noise terms of the atomic and mechanical modes, Eq. (\ref{q12}) becomes
\begin{align}
\delta\dot{a}_{m}^{r}= & -ig\delta a_{c}^{r}{\rm e}^{2i\omega_{m}t}-ig\left(\delta a_{c}^{r}\right)^{\dagger},\nonumber \\
\delta\dot{a}_{c}^{r}= & -\left[\kappa+i\left(\Delta_{c}^{\prime}+\omega_{m}\right)\right]\delta a_{c}^{r}-ig_{a}\delta c_{a}^{r}\nonumber \\
 & \ \ -ig\left(\delta a_{m}^{r}\right)^{\dagger}-ig\delta a_{m}^{r}{\rm e}^{-2i\omega_{m}t}-\sqrt{2\kappa}a_{{\rm in}}^{r},\nonumber \\
\delta\dot{c}_{a}^{r}= & -i\left(\Delta_{a}+\omega_{m}\right)\delta c_{a}^{r}-ig_{a}\delta a_{c}^{r},\label{q14}
\end{align}
in which we recognize certain terms oscillating at twice the frequency
$\omega_{m}$. When the equations are integrated over times $\omega_{m}t\gg1$,
these oscillatory terms make a negligible contribution and therefore
we may safely ignore them. After discarding the fast oscillating terms,
which is a form of the rotating-wave approximation and putting $\Delta_{c}^{\prime}=\Delta_{a}=-\omega_{m}$,
blue detuned laser pulse to both the cavity and atomic resonance,
Eq. (\ref{q14}) simplifies to 
\begin{align}
\dot{a}_{m} & =-iga_{c}^{\dagger},\nonumber \\
\dot{a}_{c} & =-\kappa a_{c}-ig_{a}c_{a}-iga_{m}^{\dagger}-\sqrt{2\kappa}a_{{\rm in}},\nonumber \\
\dot{c}_{a} & =-ig_{a}a_{c},\label{q15}
\end{align}
where, for simplicity of the notation, we have dropped $\delta$ and
the superscripts $r$ on the displacement operators.

It should be noted from Eq. (\ref{q15}) that there are both, parametric
and beam-splitter type of couplings present between the field operators.
The coupling between the cavity and mirror field operators is of the
parametric type, whereas the coupling between the cavity and atomic
field operators is of the beam-splitter type. There is no direct coupling
between the mirror and the atomic operators.

\subsection{Bad cavity limit, $\kappa\gg g,g_{a}$}

The three differential equations (\ref{q15}) can be combined into
two if we take the bad cavity limit of $\kappa\gg g_{a},g$. We can
then make an adiabatic approximation $\dot{a}_{c}\approx0$, and find
\begin{equation}
a_{c}(t)\approx-i\frac{g_{a}}{\kappa}c_{a}(t)-i\frac{g}{\kappa}a_{m}^{\dagger}(t)-\sqrt{\frac{2}{\kappa}}a_{{\rm in}}(t).\label{q16}
\end{equation}
When this result is inserted into the remaining two equations $\dot{a}_{m}$
and $\dot{c}_{a}$ in Eq. (\ref{q15}), we obtain 
\begin{align}
\dot{a}_{m} & =Ga_{m}+\sqrt{GG_{a}}c_{a}^{\dagger}+i\sqrt{2G}a_{{\rm in}}^{\dagger},\nonumber \\
\dot{c}_{a}^{\dagger} & =-G_{a}c_{a}^{\dagger}-\sqrt{GG_{a}}a_{m}-i\sqrt{2G_{a}}a_{{\rm in}}^{\dagger},\label{q17}
\end{align}
where $G=g^{2}/\kappa$ and $G_{a}=g_{a}^{2}/\kappa$. A direct integration
of Eq. (\ref{q17}) yields 
\begin{align}
a_{m}(t) & =a_{m}(0){\rm e}^{Gt}+\sqrt{GG_{a}}{\rm e}^{Gt}\int_{0}^{t}dt^{\prime}c_{a}^{\dagger}(t^{\prime}){\rm e}^{-Gt^{\prime}}\nonumber \\
 & \ \ \ \ +i\sqrt{2G}{\rm e}^{Gt}\int_{0}^{t}dt^{\prime}a_{{\rm in}}^{\dagger}(t^{\prime}){\rm e}^{-Gt},\label{q18}\\
c_{a}^{\dagger}(t) & =c_{a}^{\dagger}(0){\rm e}^{-G_{a}t}-\sqrt{GG_{a}}{\rm e}^{-G_{a}t}\int_{0}^{t}dt^{\prime}a_{m}(t^{\prime}){\rm e}^{G_{a}t^{\prime}}\nonumber \\
 & \ \ \ \ -i\sqrt{2G_{a}}{\rm e}^{-G_{a}t}\int_{0}^{t}dt^{\prime}a_{{\rm in}}^{\dagger}(t^{\prime}){\rm e}^{G_{a}t^{\prime}}.\label{q19}
\end{align}

Alternatively, we may write Eq. (\ref{q17}) in a matrix form as 
\begin{equation}
\dot{\vec{z}}(t)={\bf M}\vec{z}(t)+i\sqrt{2G}\vec{\eta}(t),\label{q20}
\end{equation}
where $\vec{z}(t)=(a_{m}(t),c_{a}^{\dagger}(t))^{T}$, the drift matrix
${\bf M}$ is given by 
\begin{align}
{\bf {M}} & =G\left(\begin{array}{cc}
1 & \sqrt{\lambda}\\
-\sqrt{\lambda} & -\lambda
\end{array}\right),\label{q21}
\end{align}
and 
\begin{equation}
\vec{\eta}(t)=\left(a_{{\rm in}}^{\dagger}(t),-\sqrt{\lambda}a_{{\rm in}}^{\dagger}(t)\right)^{T},\label{q22}
\end{equation}
with $\lambda=G_{a}/G$.

It can be seen from Eq. (\ref{q21}) that the determinant of the matrix
${\bf M}$ is zero. This means that there exists a linear combination
of the operators $a_{m}(t)$ and $c_{a}^{\dagger}(t)$ which is a
constant of motion.

It is clear from Eq. (\ref{q17}) that a linear combination 
\begin{equation}
u(t)=\sqrt{\frac{G_{a}}{|G-G_{a}|}}a_{m}(t)+\sqrt{\frac{G}{|G-G_{a}|}}c_{a}^{\dagger}(t)\label{q23}
\end{equation}
is a constant of motion, $\dot{u}(t)=0$, that $u(t)$ does not evolve
in time, $u(t)=u(0)$. However, there is an another linear combination
\begin{equation}
w(t)=\sqrt{\frac{G}{|G-G_{a}|}}a_{m}(t)+\sqrt{\frac{G_{a}}{|G-G_{a}|}}c_{a}^{\dagger}(t),\label{q24}
\end{equation}
which in the case $G_{a}\neq G$ evolves in time. It is easily verified
that the linear combinations (\ref{q23}) and (\ref{q24}) satisfy
the fundamental commutation relations 
\begin{equation}
{\color{red}|[u,u^{\dagger}]|=|[w,w^{\dagger}]|=1,}\ {\rm and}\ [u,w^{\dagger}]=0.\label{q25}
\end{equation}
We may call the linear combinations $u(t)$ and $w(t)$ as superposition
modes of the mirror and atomic modes. Since $u(t)$ is a constant
of motion, we see that the evolution of the mode $w(t)$ completely
determines the time evolution of the system.

Let us find the explicit time-dependent behavior of $w(t)$. It is
not difficult to show from Eq. (\ref{q17}) that the equation of motion
for $w(t)$ depends on whether $G>G_{a}$ or $G_{a}>G$. We will consider
separately these two cases. For the case $G>G_{a}$, we have 
\begin{equation}
\dot{w}=(G-G_{a})w+i\sqrt{2(G-G_{a})}a_{{\rm in}}^{\dagger}.\label{q26}
\end{equation}
In its time-integrated form, Eq. (\ref{q26}) is 
\begin{align}
w(t) & =w(0){\rm e}^{(G-G_{a})t}\nonumber \\
 & \ \ \ \ +i\sqrt{2(G-G_{a})}{\rm e}^{(G-G_{a})t}\int_{0}^{t}dt^{\prime}a_{{\rm in}}^{\dagger}(t^{\prime}){\rm e}^{-(G-G_{a})t^{\prime}}.\label{q27}
\end{align}

For the case $G_{a}>G$, we have 
\begin{equation}
\dot{w}=-(G_{a}-G)w-i\sqrt{2(G_{a}-G)}a_{{\rm in}}^{\dagger},\label{q28}
\end{equation}
and its time-integrated form is 
\begin{align}
w(t) & =w(0){\rm e}^{-(G_{a}-G)t}\nonumber \\
 & \ \ \ \ -i\sqrt{2(G_{a}-G)}{\rm e}^{-(G_{a}-G)t}\int_{0}^{t}dt^{\prime}a_{{\rm in}}^{\dagger}(t^{\prime}){\rm e}^{(G_{a}-G)t^{\prime}}.\label{q29}
\end{align}

We may use Eqs. (\ref{q27}) and (\ref{q29}) to evaluate the solutions
for $a_{m}(t)$ and $c_{a}(t)$. From Eqs. (\ref{q23}) and (\ref{q24})
and for $G>G_{a}$, we have by inversion 
\begin{align}
a_{m}(t) & =-\sqrt{\frac{G_{a}}{G-G_{a}}}u(0)+\sqrt{\frac{G}{G-G_{a}}}w(t),\nonumber \\
c_{a}^{\dagger}(t) & =\sqrt{\frac{G}{G-G_{a}}}u(0)-\sqrt{\frac{G_{a}}{G-G_{a}}}w(t),\label{q30}
\end{align}
which shows that the time evolution of the mirror and atomic field
modes is known once $w(t)$ has been determined from Eq. (\ref{q27}).

Similarly, for $G_{a}>G$, the solutions for $a_{m}(t)$ and $c_{a}(t)$
are 
\begin{align}
a_{m}(t) & =\sqrt{\frac{G_{a}}{G_{a}-G}}u(0)-\sqrt{\frac{G}{G_{a}-G}}w(t),\nonumber \\
c_{a}^{\dagger}(t) & =-\sqrt{\frac{G}{G_{a}-G}}u(0)+\sqrt{\frac{G_{a}}{G_{a}-G}}w(t),\label{q31}
\end{align}
where in this case $w(t)$ is given in Eq. (\ref{q29}).

We see that the solutions for $a_{m}(t)$ and $c_{a}(t)$ are easily
written in terms of $w(t)$ given in Eq. (\ref{q27}) or (\ref{q29})
depending on whether $G>G_{a}$ or $G_{a}>G$. Although the solutions
for $w(t)$ in the two cases look very similar, we will see that they
lead to quite different results for entanglement between the modes
and quantum steering.

\subsection{Input-output relations}

Having the time-dependent solutions for the field operators of the
three modes, we may now calculate relations between the input and
output amplitudes of the fields. For the cavity field, we use the
well known input and output relation 
\begin{equation}
a_{{\rm out}}^{c}(t)=a_{{\rm in}}(t)+\sqrt{2\kappa}a_{c}(t),\label{q32}
\end{equation}
which after applying Eq. (\ref{q16}) becomes 
\begin{align}
a_{{\rm out}}^{c}(t) & =-a_{{\rm in}}(t)-i\sqrt{2G_{a}}c_{a}(t)-i\sqrt{2G}a_{m}^{\dagger}(t)\nonumber \\
 & =-a_{{\rm in}}(t)-i\sqrt{2|G-G_{a}|}w^{\dagger}(t).\label{q33}
\end{align}
Since the temporal behavior of $w(t)$ depends on whether $G>G_{a}$
or $G_{a}>G$, we consider the input-output relations separately for
those two cases.

\subsubsection{1. The case $G>G_{a}$\label{Sec.2aa}}

We first consider the input-output relations in the case $G>G_{a}$.
If we insert into Eq. (\ref{q33}) the result for the adjoint of $w(t)$
from Eq. (\ref{q27}), we obtain 
\begin{align}
a_{{\rm out}}^{c}(t)= & -a_{{\rm in}}(t)-i\sqrt{2(G-G_{a})}w^{\dagger}(0){\rm e}^{(G-G_{a})t}\nonumber \\
 & -2(G-G_{a}){\rm e}^{(G-G_{a})t}\int_{0}^{t}dt^{\prime}a_{{\rm in}}(t^{\prime}){\rm e}^{-(G-G_{a})t^{\prime}}.\label{q34}
\end{align}
It is convenient to introduce a set of normalized temporal light modes
\begin{align}
A_{{\rm in}} & =\sqrt{\frac{2(G-G_{a})}{1-{\rm e}^{-2(G-G_{a})\tau}}}\int_{0}^{\tau}dt\, a_{{\rm in}}(t){\rm e}^{-(G-G_{a})t},\nonumber \\
A_{{\rm out}} & =\sqrt{\frac{2(G-G_{a})}{{\rm e}^{2(G-G_{a})\tau}-1}}\int_{0}^{\tau}dt\, a_{{\rm out}}^{c}(t){\rm e}^{(G-G_{a})t},\nonumber \\
B_{{\rm in}} & =a_{m}(0),\ B_{{\rm out}}=a_{m}(\tau),\ C_{{\rm in}}=c_{a}(0),\ C_{{\rm out}}=c_{a}(\tau),\nonumber \\
W_{{\rm in}} & =w(0),\ W_{{\rm out}}=w(\tau),\ U_{{\rm in}}=u(0),\ U_{{\rm out}}=u(\tau).\label{q35}
\end{align}
By inserting Eqs. (\ref{q30}) and (\ref{q34}) into Eq. (\ref{q35}),
we then can determine how the output field of each mode, after an
interaction time $\tau$, is related to the input fields of the modes
involved. After straightforward calculations, we find that the output
fields of the three modes are related to the input fields as 
\begin{align}
A_{{\rm out}} & =-{\rm e}^{r_{\alpha}}A_{{\rm in}}-i\alpha\sqrt{{\rm e}^{2r_{\alpha}}-1}B_{{\rm in}}^{\dagger}-i\beta\sqrt{{\rm e}^{2r_{\alpha}}-1}C_{{\rm in}},\nonumber \\
B_{{\rm out}} & =(\alpha^{2}e^{r_{\alpha}}-\beta^{2})B_{{\rm in}}+\alpha\beta\left({\rm e}^{r_{\alpha}}-1\right)C_{{\rm in}}^{\dagger}\nonumber \\
 & \ \ \ \ +i\alpha\sqrt{{\rm e}^{2r_{\alpha}}-1}A_{{\rm in}}^{\dagger},\nonumber \\
C_{{\rm out}} & =(\alpha^{2}-\beta^{2}e^{r_{\alpha}})C_{{\rm in}}-\alpha\beta\left({\rm e}^{r_{\alpha}}-1\right)B_{{\rm in}}^{\dagger}\nonumber \\
 & \ \ \ \ +i\beta\sqrt{{\rm e}^{2r_{\alpha}}-1}A_{{\rm in}}.\label{q36}
\end{align}
where 
\begin{eqnarray}
\alpha=\sqrt{\frac{G}{G-G_{a}}}, & \quad & \beta=\sqrt{\frac{G_{a}}{G-G_{a}}},\label{q37}
\end{eqnarray}
and $r_{\alpha}=(G-G_{a})\tau=G\tau/\alpha^{2}$ is the normalized
interaction time parameter.

By further expressing the $B_{{\rm i}}$ and $C_{{\rm i}}$ modes
in terms of the superposition mode $W_{{\rm i}}$ (${\rm i}=({\rm out},{\rm in})$
stands for the output and input modes), we then find, with the help
of Eqs. (\ref{q24}) and (\ref{q35}) that Eq. (\ref{q36}) becomes
\begin{align}
A_{{\rm out}} & =-{\rm e}^{r_{\alpha}}A_{{\rm in}}-i\sqrt{{\rm e}^{2r_{\alpha}}-1}W_{{\rm in}}^{\dagger},\nonumber \\
W_{{\rm out}} & ={\rm e}^{r_{\alpha}}W_{{\rm in}}+i\sqrt{{\rm e}^{2r_{\alpha}}-1}A_{{\rm in}}^{\dagger}.\label{q38}
\end{align}
A number of interesting features follow from this equation.

(i) Equation (\ref{q36}) illustrates the intrinsically two-mode behavior
of our three mode system that the cavity mode interacts effectively
with the superposition mode $w$ rather than with the mirror and atomic
modes separately.

(ii) The system transforms input fields into the output fields with
a real amplitude ${\rm e}^{r_{\alpha}}$ that for $G>G_{a}$ is greater
than one. Thus, the system might reasonably be called as an amplifier
with the gain factor ${\rm e}^{r_{\alpha}}$. It is particularly well
seen when one calculates the average number of photons in the output
modes. By using Eq. (\ref{q10}) for the correlations in the input
fields, we readily find 
\begin{align}
n_{c}(\tau) & =\langle A_{{\rm out}}^{\dagger}A_{{\rm out}}\rangle=\alpha^{2}\left(n_{0}+1\right)\left({\rm e}^{2r_{\alpha}}-1\right),\nonumber \\
n_{w}(\tau) & =\langle W_{{\rm out}}^{\dagger}W_{{\rm out}}\rangle=\alpha^{2}\left(n_{0}+1\right){\rm e}^{2r_{\alpha}}-1.\label{q39}
\end{align}
We see that the average number of photons in both modes increases
with $r$ that the output fields are amplified during the evolution
process. Note the conservation of the difference between the average
number of photons in the two modes, $n_{w}(\tau)-n_{c}(\tau)=n_{w}(0)$,
characteristic for parametric amplification.

(iii) The parameter $r_{\alpha}$ depends on the difference $G-G_{a}$
rather than the sum $G+G_{a}$ of the coupling strengths. This is
connected with the presence of two different types of couplings between
the modes, the parametric coupling between the cavity and the mirror
modes described by $G$, and the beam-splitter type coupling between
the cavity and the atomic modes described by $G_{a}$. The parametric
coupling creates squeezing (correlations) between the mirror and cavity
modes to a degree $G\tau$ that then with a degree $G_{a}\tau$ is
transferred to the atomic mode. Thus, the effective strength of the
parametric interaction between the modes is $(G-G_{a})\tau$. It is
easy to understand if one notices that the superposition mode $w$
is a combination of the operators characteristics of a two-mode squeezed
state. Noting from Eq. (\ref{q37}) that 
\begin{eqnarray}
\alpha=\cosh s, & \  & \beta=\sinh s,\label{q40}
\end{eqnarray}
where $s=arctanh\sqrt{G_{a}/G}$, it then follows from Eq. (\ref{q24})
that 
\begin{eqnarray}
w & = & a_{m}\cosh s+c_{a}^{\dagger}\sinh s=S(s)a_{m}S^{\dagger}(s).\label{q41}
\end{eqnarray}
Clearly, $w$ is the annihilation operator of a combination of two
field modes in which the annihilation operator mirror mode is couple
to the Hermitian conjugate of the atomic mode. Such a combination
is generated by the squeezing transformation of the annihilation operator
$a_{m}$ with the unitary two-mode squeeze operator 
\begin{eqnarray}
S(s) & = & {\rm e}^{s\left(a_{m}c_{a}-a_{m}^{\dagger}c_{a}^{\dagger}\right)}.\label{q42}
\end{eqnarray}
Evidently, the superposition (\ref{q41}) results from a two-mode
squeezing transformation with the two-mode squeezing parameter $s$.

(iv) The appearance of the two-mode squeezed state (\ref{q41}) is
a surprising result since according to Eq. (\ref{q15}), the mirror
and the atoms are not coupled through a parametric process. This type
of coupling is created dynamically by the fast decaying cavity mode.
The parametric coupling between the cavity and the mirror creates
a two-mode squeezed state between $a_{m}$ and $a_{c}$ modes. Due
to the fast damping of the cavity mode, the beam-splitter type coupling
between $a_{c}$ and $c_{a}$ swaps the cavity mode with the atomic
mode to form the superposition mode $w$.

(v) Equation (\ref{q40}) shows that the ratio $G_{a}/G$ determines
the degree of squeezing (correlation) between the mirror and the atoms.
When $G_{a}\rightarrow G$, the squeezing parameter $s\rightarrow\infty$,
and then the modes $a_{m}$ and $c_{a}$, which form the combination
$w$, become themselves maximally squeezed or, equivalently, maximally
entangled. Simultaneously, when $s$ increases the squeezing parameter
$r_{\alpha}$ decreases, indicating that the squeezing processes determined
by the parameters $s$ and $r_{\alpha}$ exclude each other. The simplest
way to identify the competition between these two squeezing processes
is to test the Cauchy-Schwarz inequality. Two modes are said to be
squeezed if the Cauchy-Schwarz inequality is violated. It is well
known that in the case of a Gaussian state, the Cauchy-Schwarz inequality
is violated when the anomalous cross-correlation function 
\begin{equation}
\eta=\frac{|\langle A_{{\rm out}}(\tau)W_{{\rm out}}(\tau)\rangle|^{2}}{n_{c}(\tau)n_{w}(\tau)}\label{q43}
\end{equation}
is larger than one. Using Eqs. (\ref{q36}) and (\ref{q39}), we find
\begin{equation}
\eta=\frac{(n_{0}+1){\rm e}^{2r_{\alpha}}\cosh^{2}\! s}{(n_{0}+1){\rm e}^{2r_{\alpha}}\cosh^{2}\! s-1}.\label{q44}
\end{equation}
Evidently, $\eta>1$ for any $r_{\alpha}$ and $s$ except for $s\rightarrow\infty$,
at which $\eta$ approaches no squeezing limit $(\eta=1)$ irrespective
of $r_{\alpha}$. Viewed as a function of $s$, the cross-correlation
function $\eta$ is largest for $s=0$ and decreases with an increasing
$s$. From this it follows that the modes $a_{c}$ and $w$ can be
squeezed as long as $s<\infty$ and the squeezing vanishes in the
limit of $s\rightarrow\infty$. On the other hand, at $s\rightarrow\infty$,
the modes $a_{m}$ and $c_{a}$, which form the combination $w$,
are themselves maximally squeezed. Hence, in our three-mode system
with $G>G_{a}$, squeezing can occur between the cavity mode and the
combination $w$ of the modes $a_{m}$ and $c_{a}$ if those modes
are \textit{not} themselves maximally squeezed. We shall return to
this problem in more details in Sec. \ref{sec:3} and \ref{sec:4},
where we examine entanglement and quantum steering between different
combinations of the modes.

\subsubsection{2. The case $G_{a}>G$}

A calculation similar to that of the case $G>G_{a}$ can be applied
to find the input-output relations for the case $G_{a}>G$. As in
the previous case, we determine the time evolution of the field operators
in terms of $w(t)$, which in the present case is given by Eq. (\ref{q29}).
First, we evaluate the output cavity field from the expression (\ref{q33})
and by using Eq. (\ref{q29}), we find 
\begin{align}
a_{{\rm out}}^{c}(t)= & -a_{{\rm in}}(t)-i\sqrt{2(G_{a}-G)}w^{\dagger}(0){\rm e}^{-(G_{a}-G)t}\nonumber \\
 & +2(G_{a}-G){\rm e}^{-(G_{a}-G)t}\int_{0}^{t}dt^{\prime}a_{{\rm in}}(t^{\prime}){\rm e}^{(G_{a}-G)t^{\prime}}.\label{q45:aout_Ga}
\end{align}

In analogy to the previous case, we define a set of the normalized
modes 
\begin{align}
A_{{\rm in}} & =\sqrt{\frac{2(G_{a}-G)}{{\rm e}^{2(G_{a}-G)\tau}-1}}\int_{0}^{\tau}dt\, a_{{\rm in}}(t){\rm e}^{(G_{a}-G)t},\nonumber \\
A_{{\rm out}} & =\sqrt{\frac{2(G_{a}-G)}{1-{\rm e}^{-2(G_{a}-G)\tau}}}\int_{0}^{\tau}dt\, a_{{\rm out}}^{c}(t){\rm e}^{-(G_{a}-G)t},\nonumber \\
B_{{\rm in}} & =a_{m}(0),\ B_{{\rm out}}=a_{m}(\tau),\ C_{{\rm in}}=c_{a}(0),\ C_{{\rm out}}=c_{a}(\tau),\nonumber \\
W_{{\rm in}} & =w(0),\ W_{{\rm out}}=w(\tau),\ U_{{\rm in}}=u(0),\ U_{{\rm out}}=u(\tau),\label{q46:ABC_Ga}
\end{align}
and find that the output fields of the individual modes can be expressed
in terms of the input fields as 
\begin{align}
A_{{\rm out}}= & -{\rm e}^{-r_{\beta}^{\prime}}A_{{\rm in}}-i\beta^{\prime}\sqrt{1-{\rm e}^{-2r_{\beta}^{\prime}}}B_{{\rm in}}^{\dagger}\nonumber \\
 & -i\alpha^{\prime}\sqrt{1-{\rm e}^{-2r_{\beta}^{\prime}}}C_{{\rm in}},\nonumber \\
B_{{\rm out}}= & \left(\alpha^{\prime2}-\beta^{\prime2}{\rm e}^{-r_{\beta}^{\prime}}\right)B_{{\rm in}}+\alpha^{\prime}\beta^{\prime}\left(1-{\rm e}^{-r_{\beta}^{\prime}}\right)C_{{\rm in}}^{\dagger}\nonumber \\
 & +i\beta^{\prime}\sqrt{1-{\rm e}^{-2r_{\beta}^{\prime}}}A_{{\rm in}}^{\dagger},\nonumber \\
C_{{\rm out}}= & \left(\alpha^{\prime2}{\rm e}^{-r_{\beta}^{\prime}}-\beta^{\prime2}\right)C_{{\rm in}}-\alpha^{\prime}\beta^{\prime}\left(1-{\rm e}^{-r_{\beta}^{\prime}}\right)B_{{\rm in}}^{\dagger}\nonumber \\
 & +i\alpha^{\prime}\sqrt{1-{\rm e}^{-2r_{\beta}^{\prime}}}A_{{\rm in}},\label{q47}
\end{align}
where 
\begin{eqnarray}
\alpha^{\prime}=\sqrt{\frac{G_{a}}{G_{a}-G}}, & \  & \beta^{\prime}=\sqrt{\frac{G}{G_{a}-G}},\label{q48:alp'beta'}
\end{eqnarray}
and $r_{\beta}^{\prime}=G\tau/\beta^{\prime2}=r/\beta^{\prime2}$.

By introducing the superposition mode $W_{{\rm i}}$, the input-output
relations (\ref{q47}) simplify to 
\begin{align}
A_{{\rm out}}= & -{\rm e}^{-r_{\beta}^{\prime}}A_{{\rm in}}-i\sqrt{1-{\rm e}^{-2r_{\beta}^{\prime}}}W_{{\rm in}}^{\dagger},\nonumber \\
W_{{\rm out}}= & \,{\rm e}^{-r_{\beta}^{\prime}}W_{{\rm in}}-i\sqrt{1-{\rm e}^{-2r_{\beta}^{\prime}}}A_{{\rm in}}^{\dagger}.\label{q49}
\end{align}
We see that similar to the $G>G_{a}$ case, the system effectively
behaviors as a two-mode system. However, there are two important differences
from the $G>G_{a}$ case.

(i) The system transforms input fields into the output fields with
an amplitude which falls off exponentially with the parameter $r_{\beta}^{\prime}$.
Therefore, for $G_{a}>G$ the system behaves like an attenuator. This
can be seen by considering the average number of photons in the output
modes. With the help of Eq. (\ref{q10}), the average numbers of photons
in the modes evaluated from Eq. (\ref{q49}) are 
\begin{align}
n_{c}(\tau) & =\beta^{\prime2}\left(n_{0}+1\right)\left(1-{\rm e}^{-2r_{\beta}^{\prime}}\right),\nonumber \\
n_{w}(\tau) & =\beta^{\prime2}\left(n_{0}+1\right){\rm e}^{-2r_{\beta}^{\prime}}+1.\label{q50}
\end{align}
Unlike the $G>G_{a}$ case, the number of photons in the cavity mode
increases with an increasing $r_{\beta}^{\prime}$ in expense of a
decreasing number of photons in the modes $w$ and saturates at $n_{c}(\infty)=\beta^{\prime2}\left(n_{0}+1\right)$.
Note the conservation of the sum of the number of photons, $n_{c}(\tau)+n_{w}(\tau)=n_{w}(0)$.
This implies that in the present case, the effective coupling between
the modes is of the form of beam-splitter process.

(ii) A consequence of the beam-splitter coupling between the modes
is no squeezing between the $a_{c}$ and $w$ modes. It is easy to
show. When Eqs. (\ref{q49}) and (\ref{q50}) are applied into Eq.
(\ref{q43}), we find that 
\begin{equation}
\eta=\frac{\beta^{\prime2}(n_{0}+1){\rm e}^{-2r_{\beta}^{\prime}}}{\beta^{\prime2}(n_{0}+1){\rm e}^{-2r_{\beta}^{\prime}}+1}.\label{q51}
\end{equation}
This shows that always $\eta<1$. It then follows that the Cauchy-Schwarz
inequality cannot be violated and consequently no squeezing between
the modes. However, it does not mean that squeezing or, equivalently,
entanglement cannot occur between combinations of other modes of the
system. We shall examine the possibility for entanglement between
different combinations of the modes in Sec. \ref{sec:3}.

\subsection{Quadrature components}

The information about entanglement between modes is obtained by studying
the variances of the quadrature components of the fields and their
linear combinations. To do that, we introduce the standard definitions
of the in-phase $X$ and out-of-phase $P$ quadrature components 
\begin{align}
X_{A}^{{\rm i}}=\frac{1}{\sqrt{2}}\left[A^{{\rm i}}+(A^{{\rm i}})^{\dagger}\right],\  & P_{A}^{{\rm i}}=\frac{1}{\sqrt{2}i}\left[A^{{\rm i}}-(A^{{\rm i}})^{\dagger}\right],\label{q52}
\end{align}
where $A$ denotes the annihilation operator of a particular field
mode, $A=(a_{c},a_{m},c_{a})$, and $"{\rm i}"$ stands for the output
and input modes, ${\rm i}=({\rm out,in})$.

The general expressions for the relations between the quadratures
of the input and output fields are quite lengthy, and are listed in
the Appendix. We proceed here with the relations between the quadrature
components of the cavity mode and the superposition modes $w$ and
$u$, which are convenient to study the variances of the quadrature
components and their properties.

In the case with $G>G_{a}$, we introduce quadrature components for
the input and output fields of the superposition modes 
\begin{align}
X_{w}^{{\rm i}}= & \,\alpha X_{m}^{{\rm i}}+\beta X_{c}^{{\rm i}},\quad P_{w}^{{\rm i}}=\alpha P_{m}^{{\rm i}}-\beta P_{c}^{{\rm i}},\nonumber \\
X_{u}^{{\rm i}}= & \,\beta X_{m}^{{\rm i}}+\alpha X_{c}^{{\rm i}},\quad P_{u}^{{\rm i}}=\beta P_{m}^{{\rm i}}-\alpha P_{c}^{{\rm i}},\label{q53}
\end{align}
and find that the output-input relations, Eqs. (\ref{A1}) and (\ref{A2}),
simplify to 
\begin{align}
X_{a}^{{\rm out}} & =-{\rm e}^{r_{\alpha}}X_{a}^{{\rm in}}-\sqrt{{\rm e}^{2r_{\alpha}}-1}P_{w}^{{\rm in}},\nonumber \\
X_{w}^{{\rm out}} & ={\rm e}^{r_{\alpha}}X_{w}^{{\rm in}}+\sqrt{{\rm e}^{2r_{\alpha}}-1}P_{a}^{{\rm in}},\nonumber \\
X_{u}^{{\rm out}} & =X_{u}^{{\rm in}},\label{q54}
\end{align}
and 
\begin{align}
P_{a}^{{\rm out}} & =-{\rm e}^{r_{\alpha}}P_{a}^{{\rm in}}-\sqrt{{\rm e}^{2r_{\alpha}}-1}X_{w}^{{\rm in}},\nonumber \\
P_{w}^{{\rm out}} & ={\rm e}^{r_{\alpha}}P_{w}^{{\rm in}}+\sqrt{{\rm e}^{2r_{\alpha}}-1}X_{a}^{{\rm in}},\nonumber \\
P_{u}^{{\rm out}} & =P_{u}^{{\rm in}}.\label{q55}
\end{align}

Similarly, in the case with $G_{a}>G$, by introducing quadrature
components 
\begin{align}
X_{w^{\prime}}^{{\rm i}} & =\beta^{\prime}X_{m}^{{\rm i}}+\alpha^{\prime}X_{c}^{{\rm i}},\quad P_{w^{\prime}}^{{\rm i}}=\beta^{\prime}P_{m}^{{\rm i}}-\alpha^{\prime}P_{c}^{{\rm i}},\nonumber \\
X_{u^{\prime}}^{{\rm i}} & =\alpha^{\prime}X_{m}^{{\rm i}}+\beta^{\prime}X_{c}^{{\rm i}},\quad P_{u^{\prime}}^{{\rm i}}=\alpha^{\prime}P_{m}^{{\rm i}}-\beta^{\prime}P_{c}^{{\rm i}},\label{q56}
\end{align}
we find from Eqs. (\ref{A3}) and (\ref{A4}) that 
\begin{align}
X_{a}^{{\rm out}} & =-{\rm e}^{-r_{\beta}^{\prime}}X_{a}^{{\rm in}}-\sqrt{1-{\rm e}^{-2r_{\beta}^{\prime}}}P_{w^{\prime}}^{{\rm in}},\nonumber \\
X_{w^{\prime}}^{{\rm out}} & ={\rm e}^{-r_{\beta}^{\prime}}X_{w^{\prime}}^{{\rm in}}-\sqrt{1-{\rm e}^{-2r_{\beta}^{\prime}}}P_{a}^{{\rm in}},\nonumber \\
X_{u^{\prime}}^{{\rm out}} & =X_{u^{\prime}}^{{\rm in}},\label{q57}
\end{align}
and 
\begin{align}
P_{a}^{{\rm out}}= & -{\rm e}^{-r_{\beta}^{\prime}}P_{a}^{{\rm in}}-\sqrt{1-{\rm e}^{-2r_{\beta}^{\prime}}}X_{w^{\prime}}^{{\rm in}},\nonumber \\
P_{w^{\prime}}^{{\rm out}}= & \,{\rm e}^{-r_{\beta}^{\prime}}P_{w^{\prime}}^{{\rm in}}-\sqrt{1-{\rm e}^{-2r_{\beta}^{\prime}}}X_{a}^{{\rm in}},\nonumber \\
P_{u^{\prime}}^{{\rm out}}= & \, P_{u^{\prime}}^{{\rm in}}.\label{q58}
\end{align}

Note that the $X^{{\rm out}}$ quadrature component of a given output
field depends on the $P^{{\rm in}}$ component of the other input
field, and vice versa, $P^{{\rm out}}$ component of a given output
field depends on the $X^{{\rm in}}$ component of the other field.
This clearly indicates the possibility for entanglement in the $X-P$
combination of the modes $a_{c}$ and $w$. Generally speaking, there
are three different types of possible combinations of the modes, $X-X,X-P$
and $P-P$, which can be considered when searching for entanglement
between the modes.

\subsection{Large squeezing regime, $r_{\alpha},r_{\beta}^{\prime}\gg1$.\label{sec3d}}

Before moving on to consideration of the variances of the quadrature
components, it is worthwhile to look at properties of the quadratures
in the limit of a large squeezing, $r_{\alpha},r_{\beta}^{\prime}\rightarrow\infty$.
As we have already mentioned, the presence of the $X-P$ type coupling
between the quadratures suggests a possibility for EPR correlations
between the $a_{c}$ and $w$ modes. It can be seen when one takes
the limit of a large squeezing, $r_{\alpha}\rightarrow\infty$. In
the amplification regime, $G>G_{a}$, by taking $r_{\alpha}\rightarrow\infty$
in Eqs. (\ref{q54}) and (\ref{q55}), and comparing the resulting
expressions for $X_{a}^{{\rm out}}$ and $P_{w}^{{\rm out}}$, we
find 
\begin{align}
X_{a}^{{\rm out}}= & -P_{w}^{{\rm out}}.\label{q59:super_c_m}
\end{align}
This demonstrates the possibility of a perfect EPR correlation between
the modes that independent of the state of the input fields, the $P_{w}^{{\rm out}}$
component of the mode $w$ can be predicted with certainty from a
measurement of the $X_{a}^{{\rm out}}$ component of the cavity field.

However, if we consider relations between quadrature components of
the three output fields of the system, we find 
\begin{align}
X_{a}^{{\rm out}} & =-\frac{1}{\beta}P_{c}^{{\rm out}},\quad P_{a}^{{\rm out}}=\frac{1}{\beta}X_{c}^{{\rm out}},\nonumber \\
X_{a}^{{\rm out}} & =-\frac{1}{\alpha}P_{m}^{{\rm out}},\quad P_{a}^{{\rm out}}=-\frac{1}{\alpha}X_{m}^{{\rm out}},\nonumber \\
X_{c}^{{\rm out}} & =-\frac{\beta}{\alpha}X_{m}^{{\rm out}},\quad P_{c}^{{\rm out}}=\frac{\beta}{\alpha}P_{m}^{{\rm out}}.\label{q60:XY}
\end{align}
This shows that as long as the three modes are involved $(\alpha>1,\beta\neq0)$,
only imperfect EPR correlations can be created between the output
modes at $r_{\alpha}\rightarrow\infty$. In other words, from a measurement
of one of the quadratures, for example, $X_{a}^{{\rm out}}$, we can
infer only a partial information about the quadratures $P_{c}^{{\rm out}}$
and $P_{m}^{{\rm out}}$. This is because the mirror and the atoms
are coupled to the cavity mode with unequal strengths, $G$ and $G_{a}$,
respectively. In physical terms, the unequal coupling results in a
partial distinguishability of the modes. It is interesting that even
if there are imperfect EPR correlations between any pair of the individual
modes, there exists the linear combination $w$ of the mirror and
atomic modes which exhibits a perfect EPR correlation with the cavity
mode.

The situation differs in the case of attenuation, $G_{a}>G$. In this
case, by taking the limit $r_{\beta}^{\prime}\rightarrow\infty$ in
Eqs. (\ref{q57}) and (\ref{q58}), one finds 
\begin{align}
X_{a}^{{\rm out}}= & -P_{w^{\prime}}^{{\rm in}},\quad X_{w^{\prime}}^{{\rm out}}=P_{a}^{{\rm in}},\nonumber \\
P_{a}^{{\rm out}}= & -X_{w^{\prime}}^{{\rm in}},\quad P_{w^{\prime}}^{{\rm out}}=-X_{a}^{{\rm in}}.\label{q61}
\end{align}
We see that in this case, a state transfer or, equivalently, state
swapping effect occurs between the input and output modes rather than
the creation of entanglement between the output modes. The in-phase
(out-of-phase) quadrature component of a given input field is transferred
into the out-of-phase (in-phase) component of the output field of
the other mode. Thus, in the limit of $r_{\beta}^{\prime}\rightarrow\infty$
no entanglement is created between the modes $a_{c}$ and $w$. Nevertheless,
we will demonstrate that there is still possible to create EPR type
correlations between some combinations of the modes.

\section{Bipartite entanglement\label{sec:3}}

Our interest is in the creation of entanglement, in particular, perfect
EPR entangled states between the modes of the three-mode optomechanical
system. The entanglement is associated with correlations between the
modes which are reflected in the variances in linear combinations
of the quadrature components of the modes. To quantify entanglement,
we use the DGCZ criterion \cite{DGCZ} and asymmetric criterion \cite{hr13,Simon,Tombesi}
for the variances of symmetric and asymmetric combinations of the
quadrature components of the field modes. We consider different combinations,
$X-X,X-P$ and $P-P$ of the in-phase and out-of-phase quadrature
components of the modes.

\subsection{Symmetric entanglement criteria\label{sec:3a}}

Let us first consider the DGCZ inseparability criterion for the symmetric
$X-P$ combinations of the quadrature components of the output fields.
This is quantified with the inseparability parameter 
\begin{equation}
\Delta_{i,j}=\left[\Delta\left(X_{i}^{{\rm out}}\pm P_{j}^{{\rm out}}\right)\right]^{2}+\left[\Delta\left(P_{i}^{{\rm out}}\pm X_{j}^{{\rm out}}\right)\right]^{2},\label{q62}
\end{equation}
where $i,j=a,m,c,w$, and $j\neq i$. Two modes $i$ and $j$ are
said to be entangled iff $\Delta_{i,j}<2$.

For the input fields we assume that the cavity and atomic modes are
in the ordinary vacuum state whereas the mirror field mode is in a
thermal state with occupation number $n_{0}$. Then the variances
of the input fields in the modes $w$ and $u$ are 
\begin{align}
\left(\Delta X_{w}^{{\rm in}}\right)^{2}= & \,\left(\Delta P_{w}^{{\rm in}}\right)^{2}=\alpha^{2}\left(n_{0}+\frac{1}{2}\right)+\frac{1}{2}\beta^{2},\nonumber \\
\left(\Delta X_{u}^{{\rm in}}\right)^{2}= & \,\left(\Delta P_{u}^{{\rm in}}\right)^{2}=\beta^{2}\left(n_{0}+\frac{1}{2}\right)+\frac{1}{2}\alpha^{2}.\label{eq:initial_n_s}
\end{align}

We focus first on the case $G>G_{a}$ and calculate all the possible
two-mode variances $\Delta_{i,j}$. When Eqs. (\ref{q53})-(\ref{q54})
are used in Eq. (\ref{q62}), we readily find the general expressions
for the inseparability parameter $\Delta_{i,j}$ of the symmetric
$X-P$ combinations 
\begin{align}
\Delta_{a,c}= & \,2+2\alpha^{2}\left(n_{0}+1\right)\!\left[{\rm e}^{2r_{\alpha}}-1+\beta^{2}\left(e^{r_{\alpha}}-1\right)^{2}\right],\nonumber \\
\Delta_{m,c}= & \,2\left(n_{0}+1\right)\!\left[\left(\alpha^{2}{\rm e}^{r_{\alpha}}-\beta^{2}\right)^{2}+\alpha^{2}\beta^{2}\left({\rm e}^{r_{\alpha}}-1\right)^{2}\right],\nonumber \\
\Delta_{a,m}= & \,2\left(n_{0}+1\right)\!\left[\alpha\left(\alpha e^{r_{\alpha}}-\sqrt{{\rm e}^{2r_{\alpha}}-1}\right)-\beta^{2}\right]^{2}.\label{q64}
\end{align}

Had we considered only the mirror coupled to the cavity mode, $(\alpha=1,\beta=0)$,
the parameter $\Delta_{a,m}$ would have been 
\begin{equation}
\Delta_{a,m}=2(n_{0}+1)\left({\rm e}^{r}-\sqrt{{\rm e}^{2r}-1}\right)^{2},\label{q65}
\end{equation}
with $r=G\tau$, which is the result of Hofer\textit{ et al.} \cite{hw11},
who considered EPR entanglement in a two-mode optomechanical system.

\begin{figure}[h]
\centering{}\includegraphics[width=0.9\columnwidth]{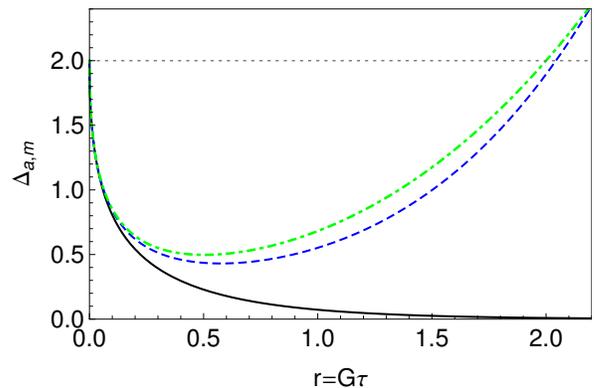} \caption{(Color online) Variation of the inseparability parameter $\Delta_{a,m}$
with $r=G\tau$ for $n_{0}=0$ and several different values of $\alpha$:
$\alpha=1$ (solid black), $\alpha=2$ (dashed blue), $\alpha=10$
(dot-dashed green).\label{fig:2} }
\end{figure}

It is easily verified from Eq. (\ref{q64}) that among the three parameters
determining bipartite correlations between the modes only $\Delta_{a,m}$
can be reduced below the separability level $2$. At $\tau=0\ (r_{\alpha}=0)$
the modes are separable and immediately afterwards, $\Delta_{a,c}$
and $\Delta_{m,c}$ begin to increase whereas $\Delta_{a,m}$ decreases
below $2$. This is shown in Fig. \ref{fig:2}, where we plot $\Delta_{a,m}$
as a function of $r$ for several different values of $\alpha$. For
$\alpha=1$, the parameter $\Delta_{a,m}$ decreases with an increasing
$r$ and at $r\rightarrow\infty$, $\Delta_{a,m}\rightarrow0$ indicating
that the state of the two modes becomes a perfect EPR state. However,
as the coupling $\alpha$ increases, $\Delta_{a,m}$ rapidly increases
at large $r$ and becomes greater than $2$. Thus, for $\alpha>1$
entanglement occurs in a restricted range of $r$ that only at small
$r$ the entanglement survives. Hence, for $\alpha>1$ the state is
not a perfect EPR state.

Although the perfect EPR state between the modes $a_{c}$ and~$a_{m}$
disappears when $\alpha>1$, it must not be thought that then there
is no possibility to create a perfect EPR state in the system. When
$\alpha>1$, the perfect EPR state is still there, but it is between
the cavity mode $a_{c}$ and the superposition mode $w$. When we
calculate the parameter $\Delta_{a,w}$, we find 
\begin{equation}
\Delta_{a,w}=2\alpha^{2}(n_{0}+1)\left({\rm e}^{r_{\alpha}}-\sqrt{{\rm e}^{2r_{\alpha}}-1}\right)^{2}.\label{q66}
\end{equation}
Apart from the appearance of the factor $\alpha^{2}$, Eq. (\ref{q66})
is formally identical with the result given in Eq. (\ref{q65}) for
the two-mode optomechanical system. The system evidently tends to
behave as a two-mode system. Therefore, $\Delta_{a,w}$ too gets reduced
below $2$ and tends to zero when $r\rightarrow\infty$. Thus, a perfect
EPR state can be created in the system even if $\alpha>1$. Since
$\alpha\geq1$, it is clear that the entanglement between $a_{c}$
and $w$ occurs in a more restricted range of $r$ than that predicted
for the two-mode optomechanical system. In the presence of the thermal
noise $(n_{0}\neq0)$, the entanglement occurs in a more restricted
area and its magnitude also drops further. It is worth noting that
the effect of the thermal noise on the parameters $\Delta_{i,j}$
is merely to increase their magnitudes by a factor $(n_{0}+1)$.

\begin{figure}[h]
\centering{}\includegraphics[width=0.9\columnwidth]{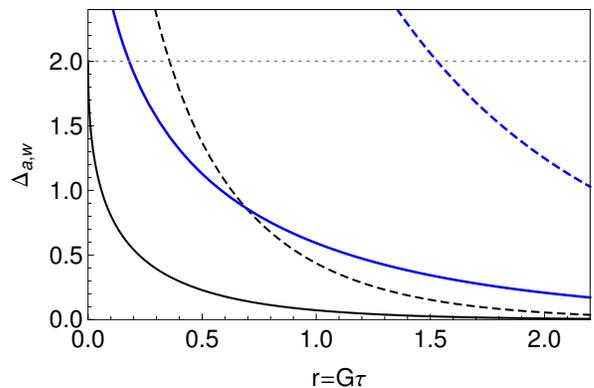} \caption{(Color online) Variation of the parameter $\Delta_{a,w}$ with $r=G\tau$
for several different values of $n_{0}$ and $\alpha$. Lower solid
black line for $n_{0}=0,\alpha=1$. Lower dashed black line for $n_{0}=5,\alpha=1$.
Upper solid blue line for $n_{0}=0,\alpha=1.5$. Upper dashed blue
line for $n_{0}=5,\alpha=1.5$. Dotted line indicates position of
$\Delta_{a,w}=2$, the threshold for entanglement.\label{fig:3} }
\end{figure}

To illustrate the behavior discussed above we show in Fig. \ref{fig:3}
the parameter $\Delta_{a,w}$ as a function of $r$ for several different
values of $n_{0}$ and $\alpha$. For $n_{0}=0$ and $\alpha=1$,
entanglement is seen to occur over the entire range of $r$ and the
created state approaches an EPR state, $\Delta_{a,w}\rightarrow0$
as $r\rightarrow\infty$. For $n_{0}>0$ entanglement occurs in the
reduced range of $r$: 
\begin{equation}
r>r_{0}=\alpha^{2}\ln\frac{\alpha^{2}(n_{0}+1)+1}{2\alpha\sqrt{n_{0}+1}}.
\end{equation}
It then follow from $r=G\tau$ that $\Delta_{a,w}<2$ everywhere except
during the short interaction time. Thus we see that the principal
effect of thermal photons and the addition of the third mode is to
add the initial noise which delays the creation of entanglement to
longer interaction times. Comparing the behavior of $\Delta_{a,w}$
with $\Delta_{a,m}$, we may conclude that with an increasing $\alpha$,
the entanglement between the modes $a_{c}$ and $a_{m}$ is transferred
to the modes $a_{c}$ and $w$. This conclusion is entirely consistent
with the conclusions reached earlier in Sec. \ref{sec3d} that for
$\alpha>1$, perfect EPR correlations are created only between the
cavity mode $a_{c}$ and the superposition mode $w$.

Turning now to the case $G_{a}>G$, we find with the help of the output-input
relations, Eqs. (\ref{q56})-(\ref{q58}), that 
\begin{align}
\Delta_{a,c}= & \,2+2\beta^{\prime2}\left(n_{0}+1\right)\!\left[1-{\rm e}^{-2r_{\beta}^{\prime}}\!+\!\alpha^{\prime2}\left(1-{\rm e}^{-r_{\beta}^{\prime}}\right)^{2}\right],\nonumber \\
\Delta_{m,c}= & \,2\!\left(n_{0}\!+\!1\right)\!\!\left[\left(\alpha^{\prime2}\!-\!\beta^{\prime2}{\rm e}^{-r_{\beta}^{\prime}}\right)^{2}\!+\!\alpha^{\prime2}\beta^{\prime2}\!\left(1\!-\!{\rm e}^{-r_{\beta}^{\prime}}\right)^{2}\right],\nonumber \\
\Delta_{a,m}= & \,2\!\left(n_{0}\!+\!1\right)\!\!\left[\alpha^{\prime2}\!-\!\beta^{\prime}\!\left(\!\beta^{\prime}{\rm e}^{-r'_{\beta}}\!+\!\sqrt{1-{\rm e}^{-2r'_{\beta}}}\,\right)\!\right]^{2},\label{q67}
\end{align}
and 
\begin{equation}
\Delta_{a,w}=2+2\beta^{\prime2}\left(n_{0}+1\right)\!\left({\rm e}^{-r'_{\beta}}+\sqrt{1-{\rm e}^{-2r'_{\beta}}}\,\right)^{2}.\label{q68}
\end{equation}

Equations (\ref{q67}) and (\ref{q68}) are markedly different from
Eqs. (\ref{q64}) and (\ref{q65}), their counterparts for the case
$G>G_{a}$. First of all, $\Delta_{a,w}$ is always greater than $2$,
and among the other parameters only $\Delta_{a,m}$ can be reduced
below $2$. Consequently, entanglement can be created only between
the cavity mode and the mirror. This indicates that in contrast to
the case $G>G_{a}$, the cavity mode entangles with the mirror alone
rather than with the mode $w$ which is the superposition of the mirror
and atomic modes.

\begin{figure}[h]
\centering{}\includegraphics[width=0.9\columnwidth]{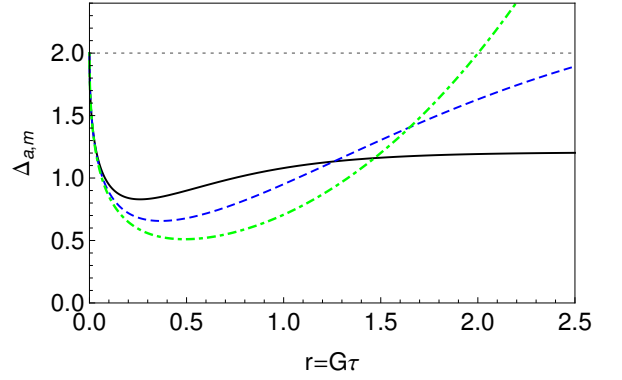} \caption{(Color online) Variation of the parameter $\Delta_{a,m}$ with $r=G\tau$
for the case $G_{a}>G$, $n_{0}=0$ and several values of $\alpha^{\prime}$:
$\alpha^{\prime}=1.2$ (solid black), $\alpha^{\prime}=1.5$ (dashed
blue), $\alpha^{\prime}=5$ (dot-dashed green).\label{fig:4} }
\end{figure}

The behavior of $\Delta_{a,m}$ given by Eq.~(\ref{q67}) is illustrated
in Fig.~\ref{fig:4} for $n_{0}=0$ and for several different values
of $\alpha^{\prime}$. For~$\alpha^{\prime}<\sqrt{2}$, which corresponds
to $G<G_{a}/2$, entanglement is seen to occur over the entire range
of $r$. For $\alpha^{\prime}>\sqrt{2}$, entanglement occurs only
in the restricted range of $r$, but it is accompanied by an enhancement
in the degree of entanglement. It is interesting that the smaller
parametric coupling strength~$G$ produces entanglement at larger
range of $r$ than the larger $G$ does. It turns out that the smallest
value of $\Delta_{a,m}$, corresponding to optimum entanglement, is
achieved when $\alpha^{\prime}\gg1$, in which case $\Delta_{a.m}=1/2$.
Hence, we may speak of $75$\% entanglement. It follows that the entanglement
is not perfect, so that the state corresponding to the maximum entanglement
is not an EPR state. 

We have already noticed an important difference between the two cases
$G>G_{a}$ and $G_{a}>G$ that in the case with $G_{a}>G$ there is
no entanglement between the cavity mode and the superposition mode
$w$. This conclusion is evident from Eq. (\ref{q68}), which clearly
shows that $\Delta_{a,w}$ cannot be reduced below $2$. This could
suggest that in the case with $G_{a}>G$, the beamsplitter type coupling
between the cavity mode and the atoms destroys the entanglement already
created between the cavity mode and the mirror. In fact, the entanglement
is not destroyed, it is still there but occurs between the mirror
and the atoms. To see this, we evaluate the separability criterion
for the $X_{m}^{{\rm out}}+X_{c}^{{\rm out}}$ and $P_{m}^{{\rm out}}-P_{c}^{{\rm out}}$
combinations of the quadrature components 
\begin{equation}
\Upsilon_{m,c}=\left[\Delta\left(X_{m}^{{\rm out}}+X_{c}^{{\rm out}}\right)\right]^{2}+\left[\Delta\left(P_{m}^{{\rm out}}-P_{c}^{{\rm out}}\right)\right]^{2}.\label{q69}
\end{equation}
The reason we evaluate variances of the $X-X$ and $P-P$ combinations
rather than that for $X-P$ combinations is in the relation between
the input-output quadrature components of the mirror and atomic fields.
According to Eqs. (\ref{A3}) and (\ref{A4}), the in-phase quadrature
components of the $a_{m}$ and $a_{c}$ modes are coupled to the in-phase
quadrature components of the input fields. The same property is seen
for the out-of-phase quadrature components. Thus, using Eqs. (\ref{A3})
and (\ref{A4}) in Eq. (\ref{q69}), we readily find 
\begin{equation}
\Upsilon_{m,c}=2\left(n_{0}+1\right)\left[1-\frac{\beta^{\prime}\left(1-{\rm e}^{-r_{\beta}^{\prime}}\right)}{\alpha^{\prime}+\beta^{\prime}}\right]^{2}.\label{q70}
\end{equation}
It is clear from Eq. (\ref{q70}) that $\Upsilon_{m,c}$ can be reduced
below $2$, but only if $r_{\beta}^{\prime}$ and $\beta^{\prime}$
are both different from zero. Since $\beta^{\prime}\neq0$ when $G\neq0$,
it follows that the presence of the parametric coupling between the
mirror and the cavity mode is necessary for entanglement between the
mirror and the atoms. We may find the minimum value of $\Upsilon_{m,c}$.
In the limit of $r\rightarrow\infty$, the parameter $\Upsilon_{m,c}$
reduces to 
\begin{equation}
\Upsilon_{m,c}=2\left(n_{0}+1\right)\left[\frac{\alpha^{\prime}}{\alpha^{\prime}+\beta^{\prime}}\right]^{2}.\label{q71}
\end{equation}
The minimum value of $\Upsilon_{m,c}$, corresponding to maximum entanglement
between the mirror and the atoms, is reached when $\alpha^{\prime},\beta^{\prime}\gg1$,
in which case $\Upsilon_{m,c}=(n_{0}+1)/2$. It follows that the maximum
$75\%$ entanglement can be achieved when~$n_{0}=0$. 
\begin{figure}[h]
\centering{}\includegraphics[width=0.9\columnwidth]{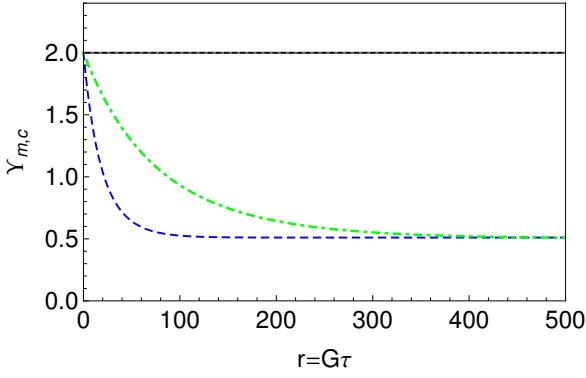} \caption{(Color online) Variation of the parameter $\Upsilon_{m,c}$ with $r=G\tau$
for the case $G_{a}>G$, $n_{0}=0$ and several values of $\alpha^{\prime}$:
$\alpha^{\prime}=1$ (solid black), $\alpha^{\prime}=5$ (dashed blue),
$\alpha^{\prime}=10$ (dot-dashed green).\label{fig:5} }
\end{figure}

The above considerations are illustrated in Fig. \ref{fig:5}, which
shows $\Upsilon_{m,c}$ as a function of $r=G\tau$ for different
values of $\alpha^{\prime}$. When $\alpha^{\prime}\neq1$, corresponding
to $G\neq0$, the entanglement is seen to occur over the entire range
of $r$. It is apparent that the entanglement increases with an increasing
$r$ and that when $\alpha^{\prime}\gg1$, the optimum entanglement
of $\Upsilon_{m,c}=1/2$ is achieved at $r\rightarrow\infty$.

\subsection{Asymmetric entanglement criteria\label{sec:3b}}

A close look at the output-input relations between the quadrature
components, Eqs. (\ref{A1})-(\ref{A4}) in the Appendix, reveals
that symmetric combinations of the output fields are accompanied by
asymmetric rather than symmetric combinations of the input fields.
These asymmetries arise not only from a difference between the coupling
strengths $G$ and $G_{a}$ but also from the presence of the thermal
noise only at the mirror. This suggests that symmetric combinations
of the output fields may not be able to detect the presence of an
entanglement between modes that might be present in an asymmetric
combination. For this reason, we now consider the criterion for asymmetric
combinations of the quadrature components \cite{hr13}, in particular,
to see if we can find an entanglement between the mirror and the atoms
which, as we have seen in Eq. (\ref{q15}), are not directly coupled
to each other.

The inseparability criterion for asymmetric $X-P$ combinations of
the quadrature components of the output modes $i$ and $j$ is confirmed
when 
\begin{align}
\left[\Delta\!\left(X_{i}^{{\rm out}}+gP_{j}^{{\rm out}}\right)\right]^{2}+\left[\Delta\!\left(P_{i}^{{\rm out}}+gX_{j}^{{\rm out}}\right)\right]^{2}<1+g^{2} & ,\label{q72}
\end{align}
where $g$ is a weight factor which is chosen to minimize the variances.
The value of $g$ which minimizes $\Delta_{a,m}^{g}$ is easily found
using the variational method. By taking the derivative of $\Delta_{a,m}^{g}$
over $g$ and setting $\partial\Delta_{a,m}^{g}/\partial g=0$, we
arrive to a quadratic equation for $g$ whose the roots can be expressed
as 
\begin{equation}
g=\frac{-b\pm\sqrt{b^{2}-4ac}}{2a},\label{q73}
\end{equation}
where $b=\left(\Delta P_{m}^{{\rm out}}\right)^{2}-\left(\Delta X_{a}^{{\rm out}}\right)^{2}$
and $c=-a=\langle X_{a}^{{\rm out}}P_{m}^{{\rm out}}\rangle$. We
then choose the root which minimizes $\Delta_{a,m}^{g}$.

A close look at Eqs. (\ref{A1})-(\ref{A4}) in the Appendix reveals
the following properties of the cross correlations between the modes
\begin{align}
\langle X_{a}^{{\rm out}}P_{c}^{{\rm out}}\rangle & =\langle P_{c}^{{\rm out}}X_{a}^{{\rm out}}\rangle\nonumber \\
 & =-\langle P_{a}^{{\rm out}}X_{c}^{{\rm out}}\rangle=-\langle X_{c}^{{\rm out}}P_{c}^{{\rm out}}\rangle,\nonumber \\
\langle X_{m}^{{\rm out}}P_{c}^{{\rm out}}\rangle & =\langle P_{c}^{{\rm out}}X_{m}^{{\rm out}}\rangle\nonumber \\
 & =\langle P_{m}^{{\rm out}}X_{c}^{{\rm out}}\rangle=\langle X_{c}^{{\rm out}}P_{m}^{{\rm out}}\rangle=0.
\end{align}
Then, it is easily verified that the left side of the inequality (\ref{q72})
for the combinations of the modes $a-c$ and $m-c$ becomes 
\begin{align}
(\Delta X_{a,m}^{out})^{2}+(\Delta P_{a,m}^{out})^{2}+g^{2}[(\Delta X_{c}^{out})^{2}+(\Delta P_{c}^{out})^{2}],
\end{align}
which is always greater than the right side $1+g^{2}$.

Entanglement is possible the asymmetric $X-P$ combinations of the
quadrature components of the modes $a$ and $m$. Further, using the
argument that the output quadrature $X_{m}^{{\rm out}}$ is related
to the input quadrature $X_{c}^{{\rm in}}$ and vice versa, the output
quadrature $X_{c}^{{\rm out}}$ is related to the input quadrature
$X_{m}^{{\rm in}}$, we will also consider the criterion involving
the $X-X$ asymmetric combination 
\begin{equation}
\Upsilon_{m,c}^{g}=\frac{\left[\Delta\left(X_{m}^{{\rm out}}+gX_{c}^{{\rm out}}\right)\right]^{2}+\left[\Delta\left(P_{m}^{{\rm out}}-gP_{c}^{{\rm out}}\right)\right]^{2}}{1+g^{2}}.\label{q74}
\end{equation}

\begin{figure}[h]
\begin{centering}
\includegraphics[width=0.8\columnwidth]{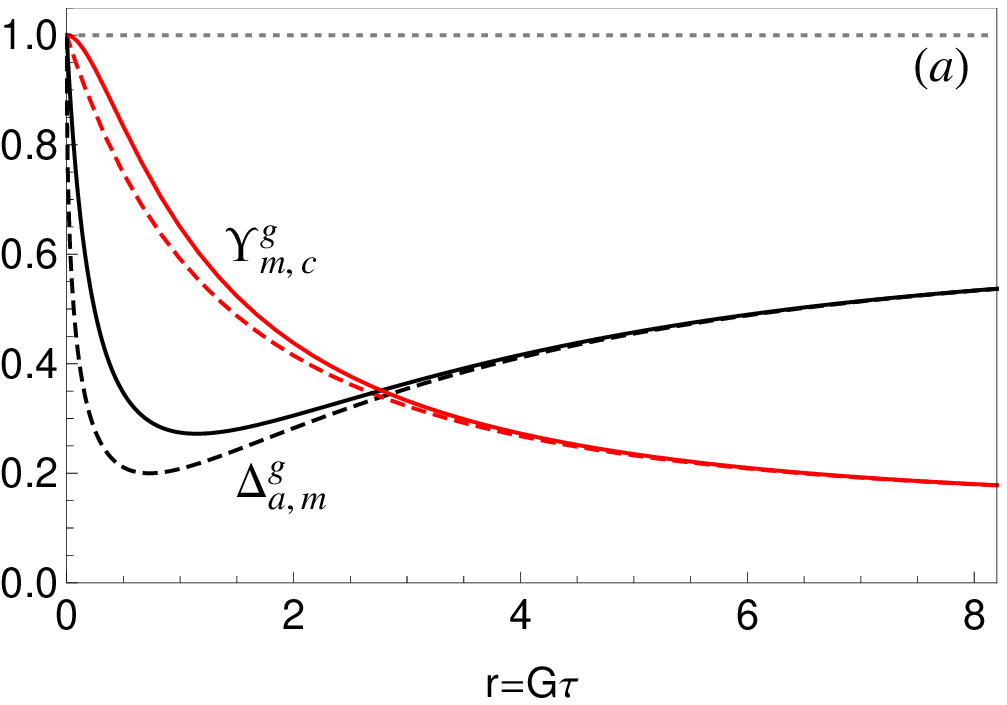} 
\par\end{centering}

\centering{}\includegraphics[width=0.8\columnwidth]{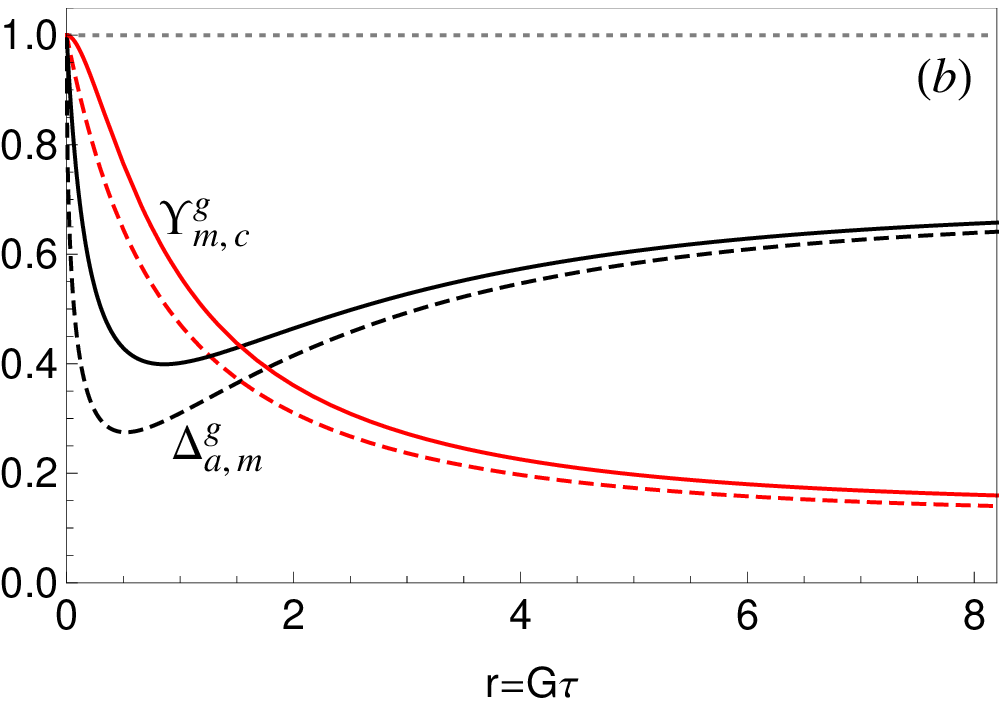} \caption{(Color online) The separability parameters for the asymmetric combinations
of the quadrature components are shown as a function of $r=G\tau$
for the case $G>G_{a}$ with $\alpha=2$ (a) and for the case $G_{a}>G$
with $\alpha^{\prime}=2$ (b). Black lines represent $\Delta_{a,m}^{g}$
and red lines represent $\Upsilon_{m,c}^{g}$, solid lines are for
$n_{0}=100$ and dashed lines for $n_{0}=0$.\label{fig:6} }
\end{figure}

The expressions for the inseparability parameters of the asymmetric
combinations of the quadratures, Eqs. (\ref{q72}), are considerably
more complex than that for the symmetric case and therefore we present
them only graphically. Some results for the cases $G>G_{a}$ and $G_{a}>G$,
and for certain combinations of the parameters $\alpha$ and $\alpha^{\prime}$,
are represented in Figs. \ref{fig:6}(a) and (b).

Figure \ref{fig:6}(a) shows the separability parameters $\Delta_{a,m}^{g}$
and $\Upsilon_{m,c}^{g}$ for the amplification case $G>G_{a}$ and
different numbers of the thermal photons $n_{0}$. Entanglement between
the mirror and the atoms as well as between the mirror and the cavity
mode is seen to occur over the entire range of $r$. At small $r$,
both $\Delta_{a,m}^{g}$ and $\Upsilon_{m,c}^{g}$ rapidly decrease
with an increasing $r$ but at larger $r$ the reduction of $\Upsilon_{m,c}^{g}$
is accompanied by a steady increase of $\Delta_{a,m}^{g}$. This clearly
demonstrate the transfer of bipartite entanglement from the pair of
modes $(a_{c},a_{m})$ to the pair $(a_{m},c_{a})$. Note that in
the limit of $r\rightarrow\infty$ and $\alpha\rightarrow\infty$,
corresponding to $G_{a}\approx G$, the parameter $\Upsilon_{m,c}^{g}$
tends to zero. This indicates that in this limit the entangled state
between the mirror and the atoms becomes a perfect EPR state. These
results are in contract to the symmetric case in which there is no
entanglement between the mirror and the atoms, and the entanglement
between the mirror and the cavity mode was restricted to small $r$.

In addition to the features mentioned above, one may note that the
separability parameters are almost insensitive to $n_{0}$ and the
sensitivity becomes increasingly unimportant as $r$ increases. This
feature is also distinctly different form that seen for the symmetric
criteria, in which the magnitudes of $\Delta_{a,m}^{g},$ $\Upsilon_{m,c}^{g}$
are enhanced by $n_{0}$ independent of $r$. Moreover, the minimum
value $r_{0}$ required for entanglement detection via the asymmetric
criteria is not limited by $n_{0}$, but in practice will depend on
the accuracy achieved for selecting the gain factors, which become
large, for the smaller $r$ values in the high $n_{0}$ limit.

Figure \ref{fig:6}(b) shows the corresponding situation for the case
$G_{a}>G$. We see that similar to the case $G>G_{a}$, entanglement
between the mirror and the atoms as well as between the mirror and
the cavity mode is seen to occur over the entire range of $r$. In
the limits of $r\rightarrow\infty$ and $\alpha^{\prime}\rightarrow\infty$,
the entangled state between the mirror and the atoms becomes a perfect
EPR state. We should mention that the results presented in Fig. \ref{fig:6}(b)
are essentially the same as that presented in Fig. \ref{fig:6}(a)
for the case $G>G_{a}$. We therefore conclude that in both cases
the asymmetric criteria predict essentially the same features for
entanglement. It then follows that the symmetric criteria cannot properly
distinguish the bipartite entanglement in the system. Despite this,
the criteria clearly demonstrate the role of the parametric coupling
between the cavity mode and the mirror in the creation of bipartite
entanglement between any other pair of the modes.

\section{Tripartite entanglement\label{sec:4}}

In the previous section we have considered bipartite entanglement
between the modes. However, the simultaneous coupling of all three
modes can result in a tripartite entanglement. In this section we
discuss how such tripartite entanglement may be generated in our three-mode
system. To see if a tripartite entanglement exists in the system we
shall consider criteria involving variances of the sums of suitably
chosen combinations of the quadrature operators of the three modes.
We shall make use of the relations between the input and output fields
given in Eqs. (\ref{A1})-(\ref{A4}), and again we discuss the cases~$G>G_{a}$
and $G_{a}>G$ separately.

In order to distinguish tripartite entanglement, we adopt the full
inseparability criterion of three modes. Within this criterion, there
is to be found two forms which involve either sums or products of
the variances of linear combinations of the quadrature operators.
We shall also consider a generalization of the full inseparability
criterion to a criterion for genuine tripartite entanglement.

It may readily be shown using Eqs. (\ref{A1})-(\ref{A4}) that the
criterion for full inseparability of our three modes requires that
any two of the following three inequalities are violated \cite{fullseparqubit,van03,olsen0507}
\begin{align}
\left[\Delta(X_{a}+P_{m})\right]^{2}+\left[\Delta(P_{a}+X_{m}+g_{c}X_{c})\right]^{2} & \geq2,\nonumber \\
\left[\Delta(X_{a}+P_{c})\right]^{2}+\left[\Delta(P_{a}+X_{c}+g_{m}X_{m})\right]^{2} & \geq2,\nonumber \\
\left[\Delta(X_{m}+X_{c})\right]^{2}+\left[\Delta(P_{m}-P_{c}+g_{a}X_{a})\right]^{2} & \geq2,\label{q75}
\end{align}
in order for the three modes to exhibit fully inseparable tripartite
entanglement. In Eq. (\ref{q75}), $g_{k}$ ($k=a,m,c$) are weight
factors which with the choice to ensure minimal values of the variances.
It should be noted that a violation of only one of the inequalities
(\ref{q75}) signals the existence of some entanglement but is not
sufficient for full inseparability.

Alternatively, one could setup a criterion involving products of the
variances instead of the sums as the set of the inequalities (\ref{q75})
can be written in a form of uncertainty principle \cite{genuineNP2012}
\begin{align}
\Delta_{am} & =\Delta(X_{a}+P_{m})\Delta(P_{a}+X_{m}+g_{c}X_{c})\geq1,\nonumber \\
\Delta_{ac} & =\Delta(X_{a}+P_{c})\Delta(P_{a}+X_{c}+g_{m}X_{m})\geq1,\nonumber \\
\Delta_{mc} & =\Delta(X_{m}+X_{c})\Delta(P_{m}-P_{c}+g_{a}X_{a})\geq1.\label{q76}
\end{align}
Each of the parameters $\Delta_{ij}$ is evaluated with the help of
the output-input relations given by Eqs. (\ref{A1})-(\ref{A4}).
The criterion (\ref{q76}) is stronger than (\ref{q75}), since if
it holds, the criterion (\ref{q75}) will also hold. Similar as for
the criterion (\ref{q75}), violation of only one of the inequalities
signals the existence of some entanglement. Violation of any two of
the inequalities demonstrates that the state is fully inseparable.

The criterion for full inseparability can be generalized to that for
the existence of genuine tripartite entanglement \cite{genuine-qubit,genuineNP2012,genuineentM}
by the requirement that 
\begin{equation}
\Delta_{{\rm sum}}=\sum\Delta_{ij}<2,\label{q77}
\end{equation}
where the sum is over all the parameters $\Delta_{ij}$. Other witnesses
of genuine tripartite entanglement have also been generalized by testing
only one inequality but with fixed gains \cite{van03,olsen0507,genuineentM}.
Here, we focus on the inequality (\ref{q76}) and then inequality
(\ref{q77}) to see whether these three modes are partially inseparable,
fully inseparable, or genuinely entangled.

\begin{figure}[t]
\begin{centering}
\includegraphics[width=0.85\columnwidth]{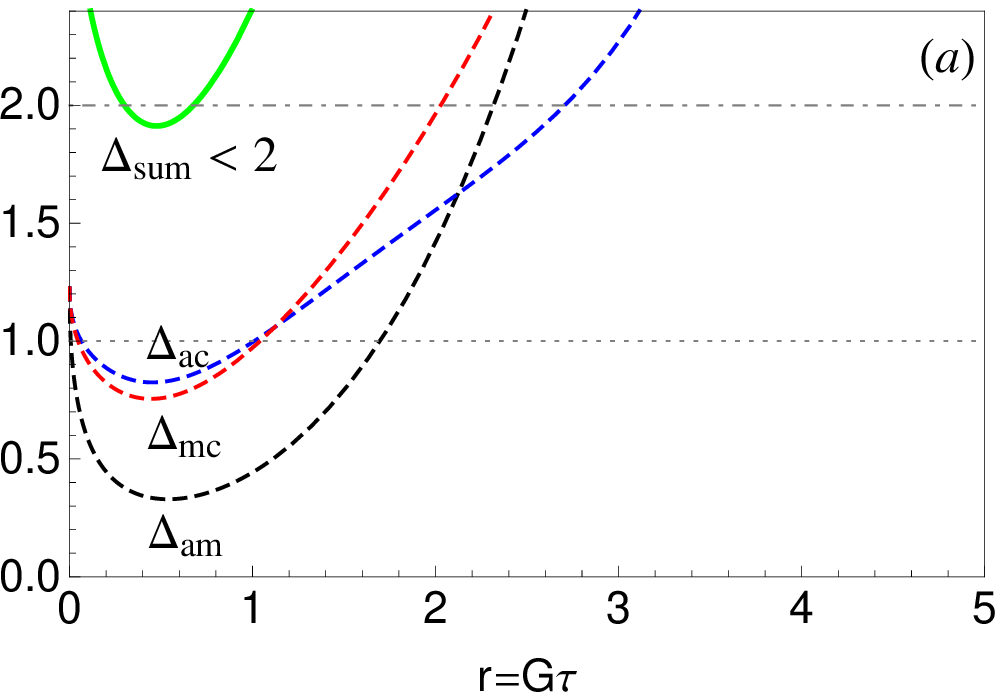} 
\par\end{centering}

\centering{}\includegraphics[width=0.85\columnwidth]{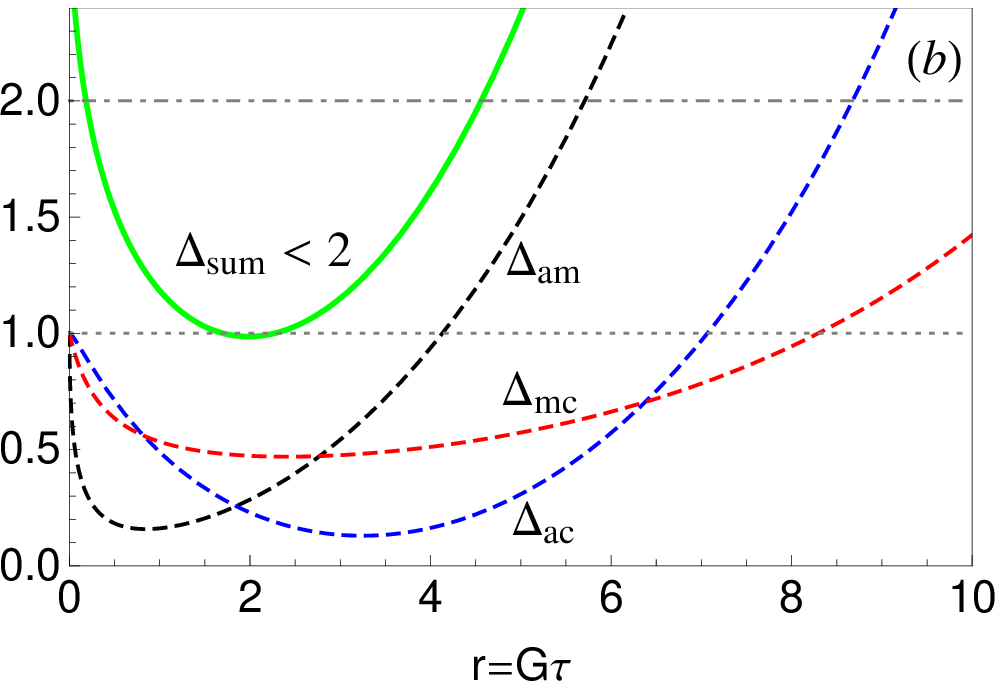}\caption{Variation of the parameters $\Delta_{ij}$ and $\Delta_{{\rm sum}}$
with $r=G\tau$ is shown for the case $G>G_{a}$ with $n_{0}=0,\alpha=2$
for (a) symmetric combinations of the quadrature operators $(g_{a}=g_{m}=g_{c}=1)$,
(b) asymmetric combinations of the quadrature operators with optimal
weight factors. The dashed line shows $\Delta_{am}$ (black), $\Delta_{ac}$
(blue), and $\Delta_{mc}$ (red). The solid green line shows $\Delta_{{\rm sum}}$.\label{fig:7} }
\end{figure}

The variation of the parameters $\Delta_{ij}$ with $r$ for the case
$G>G_{a}$ is shown in Fig. \ref{fig:7}, where frame (a) is for symmetric,
while frame (b) is for asymmetric combinations of the quadrature components.
Also shown is $\Delta_{{\rm sum}}$. We observe that in both cases
there is a range of $r$ at which two parameters of $\Delta_{am}$,
$\Delta_{ac}$, and $\Delta_{mc}$, are simultaneously less than $1$.
This means that fully inseparable tripartite entanglement can be realized
in the system. Moreover, the sum $\Delta_{{\rm sum}}$ is seen to
be less than $2$ in some range of $r$ indicating that genuine tripartite
entanglement is realized. With the minimized variances, shown in Fig.
\ref{fig:7}(b), the fully inseparable and genuine tripartite entanglements
occur over a larger range of $r$ than in the symmetric case.

The lack of the genuine tripartite entanglement at large $r$ can
be regarded to the fact that in the case $G>G_{a}$ there is a strong
tendency of the system to behave as a two-mode rather than a three-mode
system. We have seen in Sec. \ref{sec:3a} that for symmetric combinations
of the quadrature operators, a large bipartite entanglement occurs
only between the cavity and the superposition $w$ modes, see Fig.
\ref{fig:3}. The similar situation was seen for antisymmetric combinations
of the quadrature operators, illustrated in Figs. \ref{fig:6} and
\ref{fig:7}, where a large bipartite entanglement was seen only between
the mirror and atomic modes at large $r$. Thus, we may conclude that
tripartite entanglement is ruled out at the cases where a large bipartite
entanglement is present.

\begin{figure}[h]
\centering{}\includegraphics[width=0.8\columnwidth]{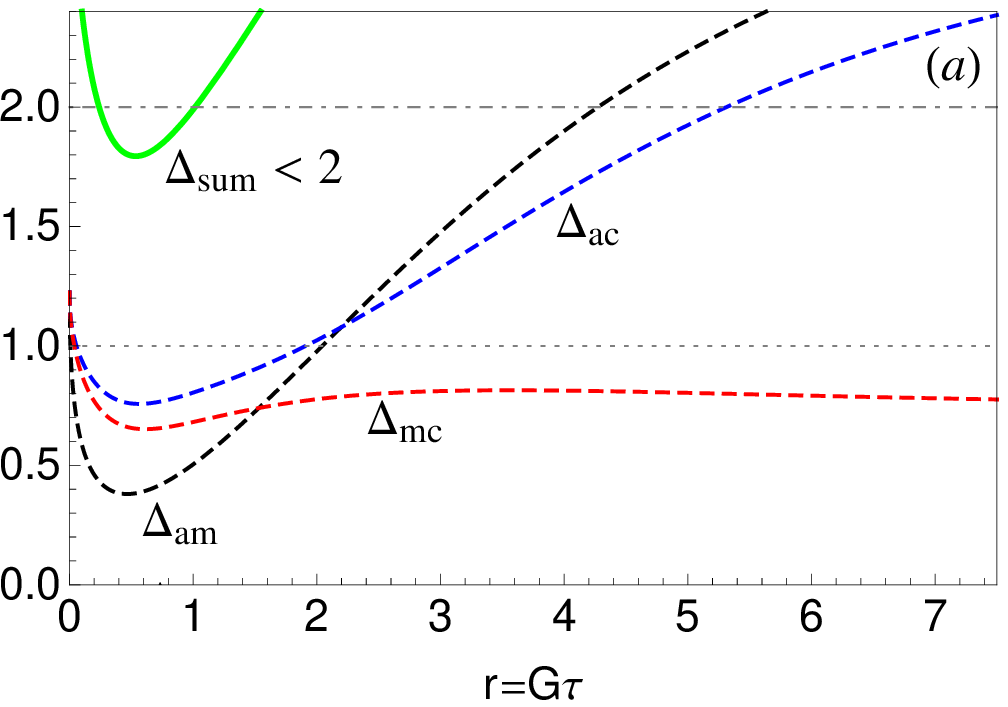}\\
 \includegraphics[width=0.8\columnwidth]{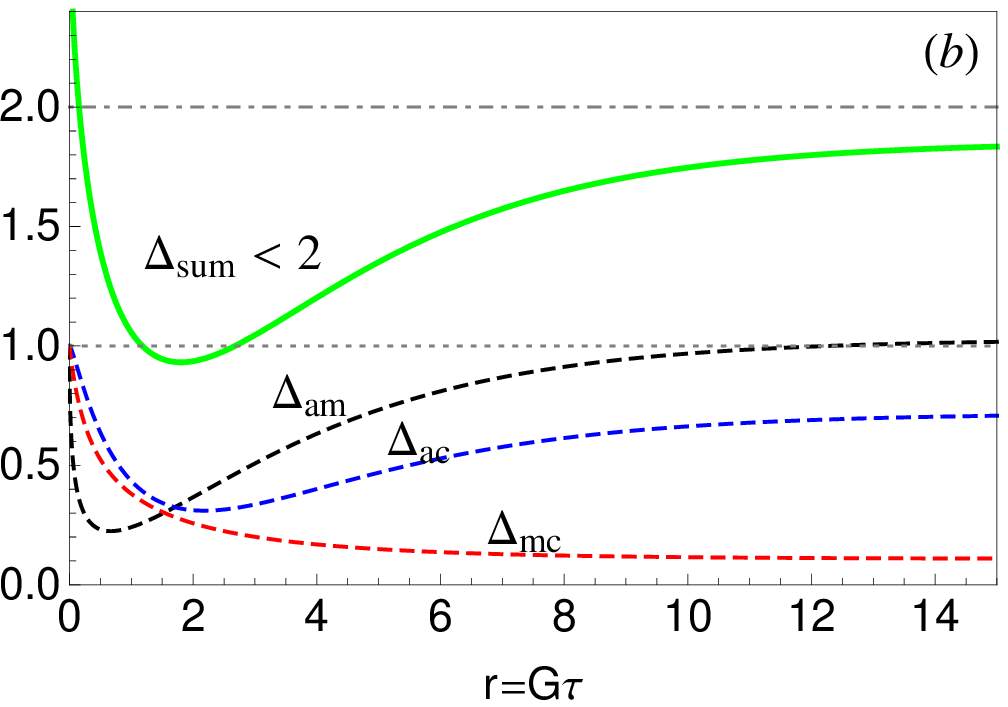} \caption{(Color online) Same as in Fig. \ref{fig:7} but for the case $G_{a}>G$
with $n_{0}=0$, $\alpha^{\prime}=2$.\label{fig:8} }
\end{figure}

Figure \ref{fig:8} shows the same situation as in Fig. \ref{fig:7}
but for $G_{a}>G$. We see that in the symmetric case there is no
a significant difference between the cases $G>G_{a}$ and $G_{a}>G$.
However, for the asymmetric case, the genuine tripartite entanglement
is present over all values of $r$.

The explanation again follows from the observation that the presence
of a tripartite entanglement is accompanied by a smaller bipartite
entanglement. In Sec. \ref{sec:3b} we saw that in the case with $G_{a}>G$
there is no entanglement between the cavity mode and the superposition
mode $w$. Thus, the entangled behaves of the system do not tend to
that of a two-mode system. The fluctuations is redistributed more
evenly between the other three pairs of modes. This resulted in a
smaller bipartite entanglement.

We may conclude that the two cases of $G>G_{a}$ and $G_{a}>G$ lead
to quite different results not only for the bipartite but also for
tripartite entanglement. It is interesting to note from Figs. \ref{fig:7}
and \ref{fig:8} that the variation of the genuine tripartite entanglement
with $r$ follows the variation of $\Delta_{a,m}$. This suggests
that the presence of an entanglement between the mirror and the cavity
mode is crucial for the genuine entanglement between the modes.

\section{Quantum steering\label{sec:5}}

We have seen in Sec. \ref{sec:3} that in the case with $G>G_{a}$,
a bipartite entanglement created between the mirror and the cavity
mode was then transferred to a pair of modes composed of the superposition
mode $w$ and the cavity mode. However, for the case with $G_{a}>G$,
the entanglement was found to be transferred to a different pair of
modes composed of the mirror and the atomic modes. It suggests that
a kind of steering behavior exists in the system that depending on
whether $G>G_{a}$ or $G_{a}>G$, the entanglement can be transferred
into different pairs of the modes. For this reason, we consider in
this section the effect of quantum steering that provides the information
how a given mode steers the other modes to be entangled. In particular,
is the steering directional? Also is it one-way or two-way steering?
Moreover, is the steering monogamic that if the mode $A$ steers $B$
then can a mode $C$ too steers $B$?

In order to develop our discussion to the problem of quantum steering,
we introduce the steering parameter defined as \cite{epr1989,howard2007,Eric2009}
\begin{equation}
E_{B|A}=\Delta_{inf,A}X_{B}\Delta_{inf,A}P_{B},\label{q81}
\end{equation}
where $\Delta_{inf,A}X_{B}\equiv\Delta\left(X_{B}|O_{A}\right)$ and
$\Delta_{inf,A}P_{B}\equiv\Delta\left(P_{B}|O'_{A}\right)$ are the
variances of the conditional distributions $P(X_{B}|O_{A})$ and $P(P_{B}|O'_{A})$,
in which $O_{A}$, $O'_{A}$ are arbitrary observables of the system
$A$, usually selected to minimize the variance product \cite{rmpMar,Eric2009}.
Quantum steering exists if 
\begin{equation}
E_{B|A}<1/2.\label{q82}
\end{equation}

Note the inherent asymmetry of the steering parameter (\ref{q81})
that $E_{B|A}<1/2$ does not necessary mean that $E_{A|B}<1/2$. We
shall refer to the situation of $E_{B|A}<1/2$ and $E_{A|B}>1/2$
as the one-way steering, and for $E_{B|A}<1/2$ and $E_{A|B}<1/2$
as a two-way steering. The asymmetry reflects the asymmetric nature
of the original EPR paradox, in which it is the reduced noise levels
of Alice's predictions for Bob's system that are relevant in establishing
the paradox \cite{EPR,onewaysteer}.

To see the quantum steering existing between the modes of our optomechanical
system we examine the conditional probabilities of the output modes
\begin{equation}
\Delta_{inf}(X_{i}^{{\rm out}}|O_{j}^{{\rm out}})=\Delta(X_{i}^{{\rm out}}+g_{j}O_{j}^{{\rm out}}),
\end{equation}
where the quadrature $O_{j}$ is selected either $O_{j}\equiv X_{j}$
or $O_{j}\equiv P_{j}$, depending on the type of the correlations
between the modes $i$ and $j$. The variances are minimized with
the choice of the weight factor 
\begin{equation}
g_{j}=-\frac{\left(\langle X_{i}^{out},O_{i}^{{\rm out}}\rangle+\langle P_{i}^{{\rm out}},O_{i}^{{\rm out}}\rangle\right)}{2\left(\Delta O_{j}^{{\rm out}}\right)^{2}}.
\end{equation}

In the following part, we give illustrative figures of the behavior
of the steering parameters $E_{j|i}$ as a function of $r$ for the
two cases $G>G_{a}$ and $G_{a}>G$. Comparison is made with the monogamy
results and the monogamy inequalities for tripartite quantum steering
recently derived by Reid \cite{genuineentM}.

\begin{figure}[h]
\centering{}\includegraphics[width=0.85\columnwidth]{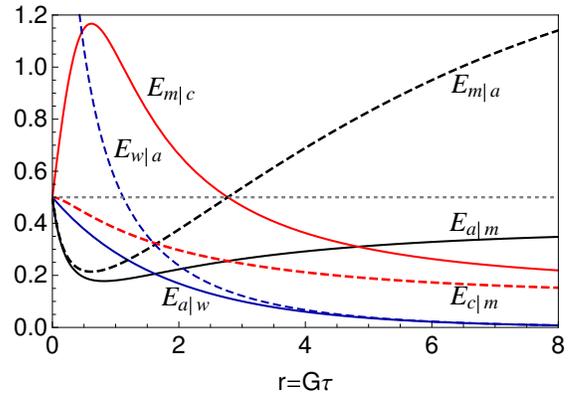} \caption{Variation of the steering parameters $E_{i|j}$ with $r$ for the
case~$G>G_{a}$, $n_{0}=0$ and $\alpha=2\,(G_{a}=0.75G)$. The solid
lines show $E_{a|m}$ (black), $E_{m|c}$ (red), and $E_{a|w}$ (blue).
The dashed lines show $E_{m|a}$ (black), $E_{c|m}$ (red), and $E_{w|a}$
(blue).\label{fig:9} }
\end{figure}

\begin{figure}[h]
\begin{centering}
\includegraphics[width=0.85\columnwidth]{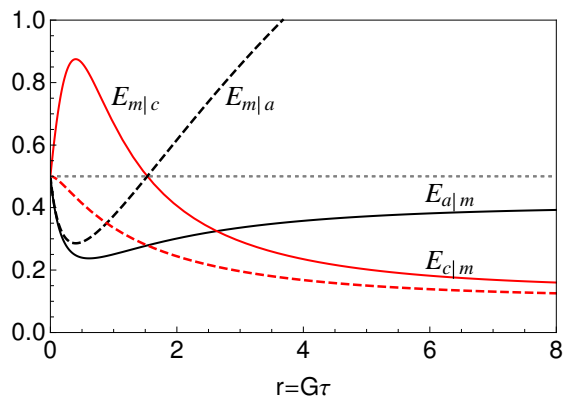} 
\par\end{centering}

\caption{Variation of the steering parameters $E_{i|j}$ with $r$ for the
case~$G_{a}>G$, $n_{0}=0$ and $\alpha^{\prime}=2\,(G_{a}=4G/3)$.
The black solid line shows $E_{a|m}$, the black dashed line $E_{m|a}$,
the red solid line $E_{m|c}$, and the red dashed line $E_{c|m}$.\label{fig:10} }
\end{figure}

Figure \ref{fig:9} shows the variation of the steering parameters
with $r$ for the case $G>G_{a}$. The figure illustrates several
interesting features, in particular, about steering monogamy and its
directionality. By inspection of the figure, we note: 
\begin{enumerate}
\item $E_{a|c}>1/2$ and $E_{c|a}>1/2$ over the entire range of $r$. Thus,
neither the cavity mode steers the atomic mode nor the atomic mode
steers the cavity mode. It is easy to understand owing to the beamsplitter
type coupling between the modes. 
\item $E_{a|w}<1/2$ over the entire range of~$r$ while $E_{w|a}<1/2$
only for $r$ greater than a some minimum value $r_{0}$. This means
that the mode $w$ always steers the cavity mode, but the cavity mode
steers $w$ in a limited range for $r$. This also means that at $r\leq r_{0}$,
one-way steering occurs between the modes, and it turns into a two-way
steering at $r>r_{0}$. This is in agreement with the result found
for quantum steering in a two-mode optomechanical system \cite{hr13}. 
\item The cavity and the mirror modes exhibit a quite different steering
behavior that $E_{a|m}<1/2$ and $E_{m|a}<1/2$ over a wide range
of $r$ but the $E_{m|a}$ steering ceases at a large $r$. It is
interested that the behavior of $E_{a|m}$ and $E_{m|a}$ is not linked
to the behavior of $E_{a|w}$ and $E_{w|a}$ but rather to the behavior
of $E_{m|c}$ and $E_{c|m}$. It is clearly seen from Fig. \ref{fig:9}
that the steering $E_{m|c}$ emerges at the same value of $r$ where
the steering $E_{m|a}$ ceases. This feature is consistent with the
monogamy result for quantum steering that two parties, the cavity
and atomic modes, cannot steer the same system, the mirror mode. 
\item According to the monogamy results of Reid \cite{genuineentM}, two
parties cannot steer the same system, but a given system can steer
two other systems. This feature is also seen in our system. It is
evident from Fig. \ref{fig:9} that $E_{a|m}<1/2$ and $E_{c|m}<1/2$
over the entire range of $r$. Thus, the dual steering is realized
that the mirror steers both the cavity mode and the atomic mode. Notice
the presence of an another dual steering that also $E_{a|m}<1/2$
and $E_{a|w}<1/2$ over the entire range of $r$. 
\item The monogamy relation of preventing the passing on of steering is
also seen in the system. Namely, it is seen from the figure that $E_{m|a}<1/2$
and $E_{c|m}<1/2$ but $E_{a|c}>1/2$. In other words, the cavity
mode steers the mirror and simultaneously the mirror steers the atomic
mode, but the atomic mode does not steer the cavity mode. 
\item A detailed inspection of the figure reveals that at small~$r$, both
$E_{a|m}<1/2$ and $E_{a|w}<1/2$. Clearly, the mirror and the mode
$w$ simultaneously steer the cavity mode. This result seems to contradict
the monogamy relation that two parties cannot steer the same system.
However, the mirror and the superposition mode $w$ are not separate
parties. The mode $w$ is a linear superposition of the mirror and
the atomic modes. In other words, the mirror mode is a part of the
mode $w$ and as such the modes cannot be treated as separate parties. 
\item It is easily verified that the monogamy inequalities $E_{a|m}E_{a|c}\geq E_{a|w}^{2}$
and $E_{a|m}+E_{a|c}\geq2E_{a|w}$ are also satisfied. However, the
inequality $E_{a|m}\geq E_{a|w}$ is violated at small $r$. It is
not difficult to see from Fig. \ref{fig:9} that for large $r$, $E_{a|m}\geq E_{a|w}$,
but for small~$r$, $E_{a|m}\geq E_{a|w}$. This discrepancy could
be understood by noting that at small $r$ the state of the system
is not in an EPR state. The inequality is satisfied at large $r$
where the state of the system approaches an EPR state. 
\end{enumerate}
Figure \ref{fig:10} shows the steering parameters for $G_{a}>G$.
We see that the dependence of the steering parameters on~$r$ is
strikingly similar to that shown in Fig. \ref{fig:9} for the case
$G>G_{a}$. Therefore, we are not going to give a detailed discussion
of the results. We just only point out that the only difference between
the two cases is that in the present case there is no steering neither
between the cavity mode and the mode $w$ nor between $w$ and the
cavity mode, i.e. $E_{a|w}>1/2$ and $E_{w|a}>1/2$ for the entire
range of $r^{\prime}$ and all the parameter's value.

In summary of this section, we have shown that the mirror is more
capable for steering of entanglement than the cavity mode which is
driven by a pulsed laser. The two way steering is found between the
mirror and the atomic ensemble despite the fact that they are not
directly coupled to each other. No quantum steering between the cavity
mode and the ensemble which are directly coupled to each other. The
reason is in the beamsplitter coupling between the modes. Thus, the
results show that there must be the parametric-type coupling present
in the system, at least between two modes.

\section{Conclusions\label{sec:6}}

We have examined entangled properties and quantum steering of a three-mode
optomechanical system composed of an atomic ensemble located inside
a single-mode cavity with a movable mirror and driven by a short laser
pulse. Using the linearization approach, we have derive analytical
expressions for the output-input relations between the amplitudes
of the fields. We have found a threshold effect for the dynamics of
the system imposed by the ratio $G/G_{a}$ of the coupling strengths
of the oscillating mirror to the cavity mode and the cavity mode to
the atoms. Above the threshold $(G>G_{a})$, the system behaves as
an amplifier, whereas below the threshold $(G<G_{a})$, the system
behaves as an attenuator of the input laser pulses. We have shown
that bipartite entanglement can be generated in both amplification
and attenuation regimes but a perfect bipartite EPR state can be generated
only in the amplification regime. The results show that in the amplification
regime the system tends to behave as a two-mode system composed of
the cavity mode and a superposition of the mirror and atomic modes.

We have also considered the inseparability criteria for tripartite
entanglement and have found that not only fully inseparable tripartite
entanglement but also genuine tripartite entanglement can be realized
in the system. The results show that in the amplification, the bipartite
and tripartite entanglements exclude each other that a large tripartite
entanglement is predicted in the range of the parameters where the
bipartite entanglement is small or even absent. The results are different
for the attenuation where tripartite entanglement occurs in the range
of the parameters where the bipartite entanglement exists between
the mirror and the cavity mode. The concept of quantum steering has
also been investigated, in particular, the ability of the system for
the directional one-way and two-way steering of entanglement. Moreover,
the monogamy relations and monogamy inequalities for quantum steering
have been analyzed. It has been found that the mirror is more capable
for steering of entanglement than the cavity mode. The two way steering
is found between the mirror and the atomic ensemble despite the fact
that they are not directly coupled to each other. The mirror can steer
both the cavity mode and the atomic mode. No quantum steering has
been found between the directly coupled cavity mode and the atomic
ensemble.

\acknowledgments This work was supported by the National Natural
Science Foundation of China under Grant No. 11274025 and 11121091.

\appendix

\section{}

In this Appendix we give the explicit expressions for the relations
between the output and input quadrature components of the fields.
If we make use of the relations between the annihilation operators
of the input and output fields and its Hermitian conjugate, Eq. (\ref{q36})
for $G>G_{a}$, and Eq. (\ref{q47}) for $G_{a}>G$, we then find
the relations between the quadrature components of the input and output
fields.

For the case $G>G_{0}$, the in-phase quadrature components satisfy
the relations 
\begin{align}
X_{a}^{{\rm out}}= & -{\rm e}^{r_{\alpha}}X_{a}^{{\rm in}}-\alpha\sqrt{{\rm e}^{2r_{\alpha}}-1}P_{m}^{{\rm in}}+\beta\sqrt{{\rm e}^{2r_{\alpha}}-1}P_{c}^{{\rm in}},\nonumber \\
X_{m}^{{\rm out}}= & \left(\alpha^{2}{\rm e}^{r_{\alpha}}-\beta^{2}\right)X_{m}^{{\rm in}}+\alpha\beta\left({\rm e}^{r_{\alpha}}-1\right)X_{c}^{{\rm in}}\nonumber \\
 & +\alpha\sqrt{{\rm e}^{2r_{\alpha}}-1}P_{a}^{{\rm in}},\nonumber \\
X_{c}^{{\rm out}}= & \left(\alpha^{2}-\beta^{2}{\rm e}^{r_{\alpha}}\right)X_{c}^{{\rm in}}-\alpha\beta\left({\rm e}^{r_{\alpha}}-1\right)X_{m}^{{\rm in}}\nonumber \\
 & -\beta\sqrt{{\rm e}^{2r_{\alpha}}-1}P_{a}^{{\rm in}},\label{A1}
\end{align}
and for the out-off-phase quadratures 
\begin{align}
P_{a}^{{\rm out}}= & -{\rm e}^{r_{\alpha}}P_{a}^{{\rm in}}-\alpha\sqrt{{\rm e}^{2r_{\alpha}}-1}X_{m}^{{\rm in}}-\beta\sqrt{{\rm e}^{2r_{\alpha}}-1}X_{c}^{{\rm in}},\nonumber \\
P_{m}^{{\rm out}}= & \left(\alpha^{2}{\rm e}^{r_{\alpha}}-\beta^{2}\right)P_{m}^{{\rm in}}-\alpha\beta\left({\rm e}^{r_{\alpha}}-1\right)P_{c}^{{\rm in}}\nonumber \\
 & +\alpha\sqrt{{\rm e}^{2r_{\alpha}}-1}X_{a}^{{\rm in}},\nonumber \\
P_{c}^{{\rm out}}= & \left(\alpha^{2}-\beta^{2}{\rm e}^{r_{\alpha}}\right)P_{c}^{{\rm in}}+\alpha\beta\left({\rm e}^{r_{\alpha}}-1\right)P_{m}^{{\rm in}}\nonumber \\
 & +\beta\sqrt{{\rm e}^{2r_{\alpha}}-1}X_{a}^{{\rm in}}.\label{A2}
\end{align}

Similarly, for the case $G_{a}>G$, we find 
\begin{align}
X_{a}^{{\rm out}}= & -{\rm e}^{-r_{\beta}^{\prime}}X_{a}^{{\rm in}}-\beta^{\prime}\sqrt{1-{\rm e}^{-2r_{\beta}^{\prime}}}P_{m}^{{\rm in}}\nonumber \\
 & +\alpha^{\prime}\sqrt{1-{\rm e}^{-2r_{\beta}^{\prime}}}P_{c}^{{\rm in}},\nonumber \\
X_{m}^{{\rm out}}= & \left(\alpha^{\prime2}-\beta^{\prime2}{\rm e}^{-r_{\beta}^{\prime}}\right)X_{m}^{{\rm in}}+\alpha^{\prime}\beta^{\prime}\left(1-{\rm e}^{-r_{\beta}^{\prime}}\right)X_{c}^{{\rm in}}\nonumber \\
 & +\beta^{\prime}\sqrt{1-{\rm e}^{-2r_{\beta}^{\prime}}}P_{a}^{{\rm in}},\nonumber \\
X_{c}^{{\rm out}}= & \left(\alpha^{\prime2}{\rm e}^{-r_{\beta}^{\prime}}-\beta^{\prime2}\right)X_{c}^{{\rm in}}-\alpha^{\prime}\beta^{\prime}\left(1-{\rm e}^{-r_{\beta}^{\prime}}\right)X_{m}^{{\rm in}}\nonumber \\
 & -\alpha^{\prime}\sqrt{1-{\rm e}^{-2r_{\beta}^{\prime}}}P_{a}^{{\rm in}},\label{A3}
\end{align}
and 
\begin{align}
P_{a}^{{\rm out}}= & -{\rm e}^{-r_{\beta}^{\prime}}P_{a}^{{\rm in}}-\beta^{\prime}\sqrt{1-{\rm e}^{-2r_{\beta}^{\prime}}}X_{m}^{{\rm in}}\nonumber \\
 & -\alpha^{\prime}\sqrt{1-{\rm e}^{-2r_{\beta}^{\prime}}}X_{c}^{{\rm in}},\nonumber \\
P_{m}^{{\rm out}}= & \left(\alpha^{\prime2}-\beta^{\prime2}{\rm e}^{-r_{\beta}^{\prime}}\right)P_{m}^{{\rm in}}-\alpha^{\prime}\beta^{\prime}\left(1-{\rm e}^{-r_{\beta}^{\prime}}\right)P_{c}^{{\rm in}}\nonumber \\
 & +\beta^{\prime}\sqrt{1-{\rm e}^{-2r}}X_{a}^{{\rm in}},\nonumber \\
P_{c}^{{\rm out}}= & \left(\alpha^{\prime2}{\rm e}^{-r_{\beta}^{\prime}}-\beta^{\prime2}\right)P_{c}^{{\rm in}}+\alpha^{\prime}\beta^{\prime}\left(1-{\rm e}^{-r_{\beta}^{\prime}}\right)P_{m}^{{\rm in}}\nonumber \\
 & +\alpha^{\prime}\sqrt{1-{\rm e}^{-2r_{\beta}^{\prime}}}X_{a}^{{\rm in}}.\label{A4}
\end{align}

\end{document}